\documentclass[twocolumn,english,aps, superscriptaddress,showpacs,showkeys]{revtex4-2}
\usepackage[utf8]{inputenc}
\usepackage[colorlinks,citecolor=blue,urlcolor=blue,linkcolor = blue]{hyperref}

\usepackage[mathscr]{euscript}

\usepackage{graphicx}
\usepackage{amsmath} 
\usepackage{amsfonts}
\usepackage{amssymb}
\usepackage{amsthm}
\usepackage{color}
\usepackage{bbold}
\usepackage{enumitem}
\usepackage{tikz}

\graphicspath{{./figs/}}

\newcommand{\rmd}{\mathrm{d}}

\DeclareMathOperator{\Tr}{Tr}

\newcommand{\nepero}{\mathrm{e}}

\newcommand{\rmua}{\mathrm{\scriptscriptstyle UA}}

\newcommand{\rmra}{\mathrm{ \scriptscriptstyle RA}}
\newcommand{\rmagp}{\mathrm{ \scriptscriptstyle AGP}}
\newcommand{\rmqa}{\mathrm{ \scriptscriptstyle QA}}

\newcommand{\rmlocalcd}{\mathrm{ \scriptscriptstyle local-CD}}

\newcommand{\ket}[1]{|{#1}\rangle}
\newcommand{\bra}[1]{\langle{#1}|}
\newcommand{\braket}[1]{\langle {#1} \rangle}

\newcommand{\zeroop}{\hat{0}}

\newcommand{\PauliSigma}{\hat{\sigma}}

\newcommand{\calS}{\mathcal{S}}

\newcommand{\Ham}{\widehat{H}}

\newcommand{\Hparam}{\widehat{H}_{\mathrm{ 0}}}

\newcommand{\Hsta}{\widehat{H}_{\mathrm{CD}}}

\newcommand{\Hra}{\widehat{H}_{\rmra}}

\newcommand{\psigs}{\psi_{\mathrm{gs}}}

\newcommand{\squaredots}{
    \vspace{-.175em}
    \tikz[line cap=round, line join=round]{
    \draw[black] (0ex,0ex) -- (0ex,0.8ex) --  (0.8ex,0.8ex) --  (0.8ex,0ex) -- cycle;
    \draw[color=black, fill=black] (0ex,0ex) circle (0.175ex);
    \draw[color=black, fill=black] (0ex,0.8ex) circle (0.175ex);
    \draw[color=black, fill=black] (0.8ex,0ex) circle (0.175ex);
    \draw[color=black, fill=black] (0.8ex,0.8ex) circle (0.175ex);
    }
}

\newcommand{\Agp}{\hat{\mathcal{A}}}

\newcommand{\Uop}{\widehat{R}}

\newcommand{\Gop}{\widehat{G}}

\newcommand{\Kop}{\widehat{\mathcal{K}}}

\newcommand{\Qop}{\widehat{\mathcal{Q}}}

\newcommand{\supmat}{{Supplementary Material}} 

\newcommand{\mainsection}[1]{{{\emph{#1--}}}}

\begin{document}

\title{ Rotated ansatz for approximate counterdiabatic driving}
\author{Glen Bigan Mbeng}
\affiliation{Universit\"at Innsbruck, Technikerstra{\ss}e 21 a, A-6020 Innsbruck, Austria}

\author{Wolfgang Lechner}
\affiliation{Universit\"at Innsbruck, Technikerstra{\ss}e 21 a, A-6020 Innsbruck, Austria}
\affiliation{Parity Quantum Computing GmbH, A-6020 Innsbruck, Austria}

\date{\today}

\begin{abstract}
    Approximate counterdiabatic (CD) protocols are a powerful tool to enhance quantum adiabatic processes that allow to reliably manipulate quantum systems on short time scales. However, implementing CD protocols entails the introduction of additional control fields in the Hamiltonian, often associated with highly non-local multi-body interactions. Here, we introduce a novel variational rotated ansatz (RA) to systematically generate experimentally accessible approximate CD protocols. We numerically benchmark our approach on state preparation and adiabatic quantum computing algorithms, and find that using RA protocols significantly enhances their performances.
\end{abstract}
\maketitle

\mainsection{Introduction}
 Adiabatic processes are a fundamental building block of quantum science and technology, with applications ranging from heat engines in thermodynamics to quantum state preparation and quantum computation \cite{Nielsen_Chuang_book2000, Vinjanampathy_Anders_ContPhys2016, Albash_Lidar_RevModPhys2018,Bohn_Rey_Sci2017}. They rely on the adiabatic theorem to manipulate quantum states by varying the system Hamiltonian at a rate proportional to the square gap between its instantaneous energy eigenvalues.
 However, despite the robustness against timing errors and thermal fluctuations,  the  efficiency  of  this  approach  is ultimately limited by the long runtimes associated with small  gaps  in  the energy spectrum. Therefore, developing and understanding shortcuts to adiabaticity (STA)~\cite{Ruschhaupt_Muga_PRL2010,Guery-Odelin_Muga_RevModPhys2019,delCampo_KimNewJPhys2019}, that mimic adiabatic dynamics with faster drivings is key to the progress of quantum technologies \cite{Bason_Morsch_NatPhys2012,Santos_Sarandy_SciRep2015,Zhou_Awschalom_NatPhys2017,Du_Zhu_PRA2017,Deng_Wu_SciAdv2018,Vepsalainen_Paraoanu_SciAdv2019}.

Counterdiabatic (CD) drivings are promising  STA  strategies that  aim at suppressing transitions between the energy levels with appropriately engineered CD Hamiltonians~\cite{Demirplak_Rice_JPhysChemA2003,Demirplak_Rice_JPhysChemB2005,Berry_JPhysA2009,Muga_Ruschhaupt_JPhysB2010,Kolodrubetz_Polkovnikov_PhysRep2017}. The exact CD Hamiltonian can mimic adiabatic dynamics using fast protocols with unit fidelity. 
However, the implementation of exact CD terms for many-body systems is experimentally challenging because it
requires a priori knowledge of the eigenstates of the Hamiltonian, and it generally involves highly non-local interactions, impossible to realize on the current hardware~\cite{delCampo_Zurek_PRL2012,Takahashi_PRE2013,Damski_2014}.

A possible strategy to generate alternative, physically feasible shortcuts is  using physical rotations and moving to an alternative Schr\"{o}dinger picture, where the non-local Hamiltonian CD terms vanish~\cite{Ibanez_Muga_PRL2012,Ibanez_Muga_PRA2015}. On the one hand, Refs.~\cite{Martinez-Garaot_Muga_PRA2014,Torrontegui_Muga_PRA2014,Kang_Xia_PRA2018,Gungordu_Nakahara_PRA2012,Takahashi_PRA2015,Bukov_Polkovnikov_PhysRevX2019} used an approach based on Lie Algebras to compute necessary physical rotation or few-level systems. On the other hand, Refs.~\cite{Deffner_delCampo_PRX2014,Sels_Polkovnikov_PNAS2017,Agundez_Blaauboer_PRA2017,delCampo_PRL2013} developed problem-specific methods to generate physical rotation for specific many-body Hamiltonians. However, a systematic approach to determining the physical unitary transformations in  many-body systems is not available.  Recently, 
Ref.~\cite{Sels_Polkovnikov_PNAS2017} introduced a variational approach to finding approximate CD driving Hamiltonians for many-body systems. This method has been recently applied both theoretically~\cite{Hartmann_Lechner_NJPhys2019,Claeys_Polkovnikov_PRL2019,Passarelli_PRR2020,Hartmann_Lechner_PRR2020,Hartmann_Lechner_Quantum2020,Villazon_Chandran_PRB2021,Hartmann_Lechner_PRA2022} and experimentally \cite{Zhou_Peng_PhysRevApp2020,Narendra_Chen_PhysRevApp2021} to speed up adiabatic processes in quantum spin systems.
Our work aims at connecting the idea of multiple Schr\"{o}dinger pictures presented in Ref.~\cite{Ibanez_Muga_PRL2012} with the variational approach of Ref.~\cite{Sels_Polkovnikov_PNAS2017}.

Here, we introduce the \textit{rotated ansatz} method to  simultaneously determine optimal approximate CD terms and the associated physical rotation. The resulting protocol, which can be computed efficiently for various many-body systems, does not require additional terms in the Hamiltonian.
After introducing the novel variational approach to determine RA protocols, we study the efficacy of the protocols generated for (i) a two-level system, (ii) a non-integrable Ising chain and (iii) a fully programmable quantum annealing architecture~\cite{Lechner_Sci2015}. We find significant performance enhancements for the protocols in all three cases.

\mainsection{Rotated ansatz.}
We consider the parametric Hamiltonian $\Hparam(\lambda)$, which depends on the real control
parameter $\lambda$, with instantaneous eigenvalues and eigenvectors $\epsilon_m(\lambda)$ and $\ket{\epsilon_m(\lambda)}$. Given an initial value $\lambda_i$ and a final value $\lambda_f$, we focus on the task of driving the system from $\ket{\epsilon_m(\lambda_i)}$ to $\ket{\epsilon_m(\lambda_f)}$ in a time $\tau$.
According to the adiabatic theorem, if $\lambda(t)$ ramps sufficiently slowly from $\lambda(0)=\lambda_i$ to $\lambda(\tau)=\lambda_f$, the unassisted (UA) driving Hamiltonian $\Ham_{\rmua}(t)=\Hparam(\lambda(t))$ implements the desired transformation~\cite{Messiah_Book2014,Jansen_Seiler_JMatPhys2007}. 
However, in practical cases, the finite coherent evolution time $\tau$ introduces diabatic transitions, leading to a fidelity loss.

The CD method aims at suppressing diabatic transitions by adding auxiliary terms to the driving Hamiltonian:
\begin{equation}\label{eqn:CD_Hamiltonian}
    \Hsta (t) = \Hparam(\lambda) + \dot{\lambda} \Agp\;.
\end{equation}
When $\Agp$ coincides with the (exact) adiabatic gauge potential (AGP) \cite{Berry_JPhysA2009,Kolodrubetz_Polkovnikov_PhysRep2017}:
{
\small
\begin{equation}\label{eqn:exact-AGP}
    \Agp^{\rmagp}
    = i\hbar \sum_{m \ne l} \frac{\bra{\epsilon_m(\lambda)}  \partial_\lambda \Ham_0(\lambda) \ket{ \epsilon_l(\lambda)}}{\epsilon_l(\lambda) - \epsilon_m(\lambda)} 
			\ket{\epsilon_m(\lambda)}\bra{\epsilon_l(\lambda)}
		\;,
\end{equation}
}%
the Schr\"{o}dinger dynamics $i\hbar\partial_t\ket{\psi(t)} = \Hsta\ket{\psi(t)}$ results in a perfect driving, that evolves $\ket{\psi(0)}=\ket{\epsilon(\lambda_i)}$ into $\ket{\psi(\tau)}=\ket{\epsilon(\lambda_f)}$. 

 The direct computation of the exact AGP in non-integrable many-body systems has an exponential complexity cost due to the lack of knowledge of spectral information \cite{Cubitt_Nat2015}. Moreover, the exact AGP can be highly non-local and, thus, hard to implement in experiments involving many-body systems. We use the least action principle introduced in Ref.~\cite{Sels_Polkovnikov_PNAS2017} to overcome these limitations. The approach is based on the observation that the exact AGP of Eq.~\eqref{eqn:exact-AGP} minimizes the following action \cite{Sels_Polkovnikov_PNAS2017,Kolodrubetz_Polkovnikov_PhysRep2017}:
\begin{equation}\label{eqn:CD_action}
\calS (\Agp)
=\Tr[\Gop_{\lambda}^2]\,, \qquad \Gop_\lambda \equiv 
\partial_\lambda \Hparam - \frac{i}{\hbar} [\Hparam, \Agp]\;.\;
\end{equation}
Therefore, we can find approximate AGPs by minimizing the action over a specific variational ansatz for $\Agp$. In particular,  restricting the ansatz to the subspace of Hamiltonian terms available in the lab generates experimentally accessible AGPs for many-body systems. 

As the AGP is approximated, the resulting CD protocol is usually associated with a finite final fidelity. In this cases, improving the effectiveness of the CD driving requires additional Hamiltonian terms and control fields. However, we can alternatively implement the CD driving in a rotating frame. Following Refs.~\cite{Ibanez_Muga_PRL2012,Takahashi_PRA2015}, we introduce a unitary transformation $\ket{\tilde{\psi}(t)} = \Uop(t)\ket{\psi(t)}$. In the new frame, the Schr\"{o}dinger equation describing the CD driving reads 
\begin{eqnarray}
    i\hbar\frac{\rmd}{\rmd t}\ket{\tilde{\psi}(t)} = \Hra(t)\ket{\tilde{\psi}(t)}\;,
\end{eqnarray}
where $\Hra(t)=\Uop\Hsta \Uop^\dagger - i\hbar \Uop\dot{\Uop}^\dagger$. Using Eq.~\eqref{eqn:CD_Hamiltonian}, we express $\Agp$ as function of $\Hra$ and $\Uop$:
\begin{eqnarray}\label{eqn:Ara-expression}
   \dot{\lambda}\Agp =\Hsta - \Ham_0
   = \Uop^\dagger\left(\Hra +i\hbar \Uop\dot{\Uop}^\dagger\right) \Uop - \Ham_0\;.
\end{eqnarray}
Next, we simultaneously determine which $\Hra(t)$ and $\Uop(t)$  result in an optimal approximate AGP $\Agp$. Since Eq.~\eqref{eqn:Ara-expression} simultaneously  depends on $\Uop$ and $\dot{\Uop}$, we cannot yet apply a time-local least action principle. To solve this issue, we introduce the auxiliary potential $\Kop = \Hra+i\hbar \Uop\dot{\Uop}^\dagger - \Ham_0 $. Furthermore, we simplify the algebra by considering only abelian transformations $\Uop(t)=\nepero^{ -\frac{i}{\hbar} \Qop(t) } $, where the generator $\Qop(t)$ commutes at different times ($[\Qop(t), \Qop(t')]=\zeroop$). We then get
\begin{align}
    \dot{\lambda}\Agp
    &=     \nepero^{ \frac{i}{\hbar} \Qop } 
\left(\Ham_0 + \Kop\right)     \nepero^{ -\frac{i}{\hbar} \Qop } 
 - \Ham_0\label{eqn:ra_AGP_definition}\\
    \Hra
    &=\Hparam+\Kop + \dot{\Qop} \label{eqn:Hra_definition}\;.
\end{align}
We interpret Eq.~\eqref{eqn:ra_AGP_definition} as a variational rotated ansatz (RA) for the AGP, which fundamentally differs from previous proposals~\cite{Sels_Polkovnikov_PNAS2017,Claeys_Polkovnikov_PRL2019,Passarelli_PRR2020}. Our ansatz  
depends on the derivative $\dot{\lambda}$, however, we can still find the optimal values of the RA auxiliary potentials $\Kop$ and $\Qop$, minimizing the action $\calS(\Agp)$ over the subspace of allowed physical operators. Then, Eq.~\eqref{eqn:Hra_definition} guarantees that the resulting driving Hamiltonian $\Hra$ also belongs to the subspace of allowed operators.

The Hamiltonian $\Hra(t)$ drives the rotated state $\ket{\tilde{\psi}(t)}$. In particular, if time boundary conditions
\begin{align}\label{eqn:ra_boundary_conditions}
    |\braket{\epsilon(\lambda)|\tilde{\psi}(t)}|&=|\braket{\epsilon(\lambda)|\psi(t)}|
    \qquad \mathrm{for}\; t = 0,\tau
\end{align}
hold, $\Hra(t)$ implements an alternative approximate STA. In the \supmat, we show that we can enforce such boundary conditions by using smooth ramps such that $\dot{\lambda}(0)=\dot{\lambda}(\tau)=0$. In this case, the new terms appearing in Eq.~\eqref{eqn:Hra_definition} enhance the performance of the RA driving with respect to the conventional unassisted driving. 

In most applications, we must rely on a numerical method to minimize the action and find $\Hra(t)$ at each time. In the \supmat{}, we describe a sequential local minimization procedure to carry out this task. The bottleneck of the algorithm is the evaluation of the action. However, in the following, we explicitly show that the complexity cost of computing the RA action is polynomial in the system size $N$, for various relevant Hamiltonians.

\mainsection{Applications} \label{sec:Applications}
We validate the RA approach on three problems: a two-level system, a quantum Ising chain, and quantum annealing. 
For each problem, we simultaneously compare the unassisted (UA) driving  $\Ham_0(t)$, the rotated ansatz (RA) $\Ham_{\rmra}(t)$ and the purely imaginary local CD driving%
~\cite{Sels_Polkovnikov_PNAS2017,Hartmann_Lechner_NJPhys2019,Prielinger_Nishimori_PRR2021}\footnote{The AGP of models with time-reversal symmetry (real symmetric Hamiltonians) are purely immagianry~\cite{Sels_Polkovnikov_PNAS2017}. Since we consider only time-reversal symmetric Hamiltonians, Eq.~\eqref{eqn:Agp_local_CD} describes the most general local approximate AGP.}%
:
 \begin{equation}\label{eqn:Agp_CD1}
    \Ham_{\rmlocalcd}(t)=\Ham_0(\lambda) + \dot{\lambda}\Agp\,, \qquad \Agp=\sum_{j=1}^N\alpha_j(\lambda) \PauliSigma_j^y\;,
 \end{equation}
where the coefficients $\alpha_j(\lambda)$ minimize the action. The three protocols do not require additional non-local control fields. However, the local CD protocol which was extensively studied in Refs.~\cite{Sels_Polkovnikov_PNAS2017, Hartmann_Lechner_NJPhys2019}, introduces new local control fields.

In all examples, we use natural units and set $\hbar=1$. We, then, consider the following smooth ramp \cite{Sels_Polkovnikov_PNAS2017, Passarelli_PRR2020, Hartmann_Lechner_NJPhys2019,Prielinger_Nishimori_PRR2021,Claeys_Polkovnikov_PRL2019} $
\lambda(t)=\sin^2\left[\frac{\pi}{2}\sin^2\left(\frac{\pi t}{2\tau} \right)\right]$,
going from $\lambda_i=0$ to $\lambda_f=1$. Here, the derivatives $\dot{\lambda}$ and $\ddot{\lambda}$ vanish at the beginning and end of
the protocol, ensuring ensuring the validity of Eq.~\eqref{eqn:ra_boundary_conditions}.
We initialize the system in $\Ham_0(\lambda_i)$'s ground state and calculate the fidelity between the time-evolved state $\ket{\psi(t)}$ and the instantaneous ground state $\ket{\epsilon_0(\lambda)}$ of $\Ham_0(\lambda)$:
\begin{eqnarray}
   F(t)=\big|\braket{\epsilon_0(\lambda)| \psi(t)}\big|^2\;.
\end{eqnarray}
The final fidelity $0\leq F(\tau)\leq 1$ measures the probability of successfully preparing $\Ham_0(\lambda_f)$'s ground state. To account for the physical rotation $\Uop(t)$, for the RA protocol we also compute the fidelity between $\ket{\psi(t)}$ and the rotated ground state $\ket{\tilde{\epsilon}_0(\lambda)}=\Uop(t)\ket{\epsilon_0(\lambda)}$:
\begin{eqnarray}\label{eqn:Agp_local_CD}
   \tilde{F}(t)=\big|\braket{\tilde{\epsilon}_0(\lambda)| \psi(t)}\big|^2 = \big|\bra{\epsilon_0(\lambda)}\Uop^\dagger(t) \ket{ \psi(t)}\big|^2\;.
\end{eqnarray}
The boundary conditions in Eq.~\eqref{eqn:ra_boundary_conditions} imply that $\tilde{F}(0)=F(0)=1$ and $\tilde{F}(\tau)=F(\tau)$. However, for intermediate times, we can have $\tilde{F}(t)\neq F(t)$. 

\mainsection{(i) Two-level system (Bell state).}\label{subsec:2LevelSys}
To illustrate the RA approach in a simple context, we consider the following two spin system~\cite{Petiziol_PRA2018,Claeys_Polkovnikov_PRL2019}:
\begin{equation}\label{eqn:2spin_H0}
    \Ham_{\mathrm{two-spins}} = -h(t)\, (\PauliSigma_1^z + \PauliSigma_2^z) +  J(t)\,(\PauliSigma_1^x\PauliSigma_2^x + \PauliSigma_1^z\PauliSigma_2^z) \,,
\end{equation}
where $J(t)$ and $h (t)$ are the control fields.
Eq.~\eqref{eqn:2spin_H0} describes a two-level system where $\Ham(t)$
only couples the states $\ket{\uparrow\uparrow}$ and $\ket{\downarrow\downarrow}$. This observation allowed Refs.~\cite{Petiziol_PRA2018,Claeys_Polkovnikov_PRL2019} to engineer an effective CD driving for this problem based on fast oscillating fields. Here, we use the system's two-level nature to analytically compute the RA driving Hamiltonian. 

We focus on the UA protocol $J_0(t)=-1$, $h_0(t)=5(1-\lambda)$, which aims at preparing the maximally entangled Bell state $\ket{\psigs(\lambda_f)}=\frac{\ket{\uparrow\uparrow} + \ket{\downarrow\downarrow}}{\sqrt{2}}$. We parametrize the auxiliary potentials as $\Qop=-\gamma\,(\PauliSigma_1^z + \PauliSigma_2^z)$ and $\Kop=\beta\,(\PauliSigma_1^x\PauliSigma_2^x + \PauliSigma_1^z\PauliSigma_2^z)$. In the \supmat{} we compute and minimize the action for the two spin problem. The variational minimization leads to the following optimal functions:
 \begin{align}\label{eqn:2level_tbeta_opt}
    \beta(t) = \frac{1}{\dot{\lambda}}\sqrt{J_0^2 +  \varphi_0^2}-J_0 \,,\;
    \gamma(t) = \frac{1}{4} \arctan\left( \frac{\varphi_0}{J_0}\right)\,,
\end{align}
 with $\varphi_0(t)=\frac{  \dot{J}_0 h_0 -J_0 \dot{h}_0}{J_0^2+4 h_0^2}$.
 The resulting analytical expression for the RA driving functions $J_{\rmra}(t)=J (t)+\dot{\gamma}(t)$, $h_{\rmra}(t)=h(t)-\beta(t)$, are a specific feature of two level systems.  In particular, this RA driving implements the exact AGP, which prepares the bell state with unitary fidelity $F_{\rmra}(\tau)=1$ for any protocol duration $\tau$. In the \supmat, we verified that our sequential minimization algorithm correctly reproduces these analytical results.

\mainsection{(ii)Non-integrable spin chain.}\label{sec:ising_chain}
\begin{figure}
    \centering
    \includegraphics[width=8.5cm]{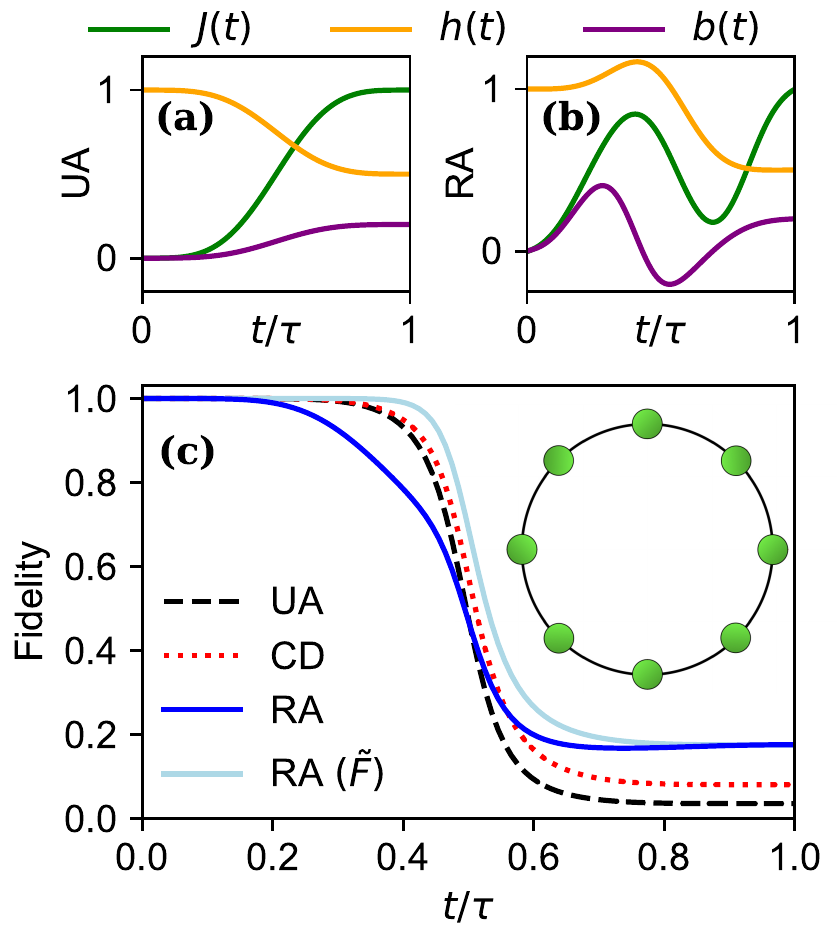}
    \caption{State preparation on a quantum Ising chain. (a) and (b) control fields for the UA and RA  protocols. (c) ground state fidelity $F(t)$ for the UA, local CD and RA protocols, and rotated ground state fidelity $\tilde{F}(t)$ for the RA protocol.  The data refers to a transitionally invariant chain of $N=8$ spins with periodic boundary conditions and protocols of duration $\tau=1$. The inset of (c) shows a cartoon spin chain.}
    \label{fig:IsingChain}
\end{figure}
Next, we use the RA approach to improve the state preparation protocols in a non-integrable quantum many-body systems. We consider a chain of $N$ spin-$\frac{1}{2}$, with the following drving Hamiltonian:
\begin{equation}\label{eqn:1DIsing_H0}
   \Ham_{\mathrm{spin-chain}} = -\sum_{j=1}^{N}\left(J (t)\PauliSigma_{j}^z\PauliSigma_{j+1}^z +b(t)\PauliSigma^z_j+h(t)\PauliSigma^x_j\right)\;, 
\end{equation}
where we assume periodic boundary conditions ($\PauliSigma_{N+1}^z\equiv\PauliSigma_{1}^z$).
Equation~\eqref{eqn:1DIsing_H0} describes the nearest neighbor quantum Ising chain with transverse and longitudinal fields. It can be experimentally realized
with superconducting qubits~\cite{Barends_Nat2016}, cold atoms~\cite{Simon_Nat2011}, and trapped ions~\cite{Jurcevic_Nat2014}.

We consider the unassisted driving $J_0(t)=\lambda$, $b_0(t)=\lambda/5$ and $h_0 (t)=1-\lambda/2$  . This path connects the initial product state $\ket{\psigs(\lambda_i)} = \ket{+}^{\otimes N}$ with the non-trivial groundstate of Eq.~\eqref{eqn:1DIsing_H0} at $h_f=1/2, J_f=1, b_f=1/5$. 
We parameterize the auxiliary potentials as $\Qop=\gamma\sum_{j}\PauliSigma_j^z\PauliSigma_{j+1}^z+ \phi \sum_{j}\PauliSigma_j^z$ and $\Kop=\beta\sum_{j}\PauliSigma_j^x$. 
Then, the RA driving functions read $J_{\rmra}=J_0 -\dot{\gamma}$, $b_{\rmra}=b_0-\dot{\phi}$ and $h_{\rmra}=h_0-\beta$. 

Due to the translational symmetry and the short range of the interaction,  the ratio $\calS/( N2^{N})$ is independent of system size $N$ (for $N\geq 4
$), as we prove in the \supmat. Thus, we can readily 
compute $\calS$  by considering a system of $N=4$ spins. We provide the resulting analytical expression of the action in the \supmat. 

We benchmark the RA method on a protocol of duration $\tau=1$. Using the sequential action minimization we find the optimal functions $J^{\rmra}(t)$, $b^{\rmra}(t)$ and $h^{\rmra}(t)$ [Fig.~\ref{fig:IsingChain}(b)], which do not depend on $N$. In Fig.~\ref{fig:IsingChain}(c), we compare instantaneous ground state fidelity of the UA (dashed black line), the local CD (dotted red line) and the RA (solid blue line) protocols, on a chain of $N=8$ spins. The diabatic transitions in the UA dynamics  gradually reduce the instantaneous fidelity $F_{\rmua}(t)$ (black dashed line), leading to the final value $F_{\rmua}(\tau)=8\%$. The local CD driving enhances the fidelity at all times $F_{\rmlocalcd}(t)\geq F_{\rmua}(t)$, leading to a greater final fidelity $F_{\rmlocalcd}(\tau)=18\%$. The RA's fidelity $F_{\rmra}(t)$, is lower than the UA's fidelity for intermediate times. However, the RA protocol actually boosts the final fidelity up to $F_{\rmra}(\tau)=36\%$. Indeed, the implemented RA driving, instantaneously tracks the rotated ground state with a fidelity $\tilde{F}_{\rmra}(t)$ (light blue dot-dashed line) which is greater than those achieved by the UA and the local CD drivings.

\mainsection{(iii) Quantum annealing.}
\begin{figure}
    \centering
    \includegraphics[width=8.5cm]{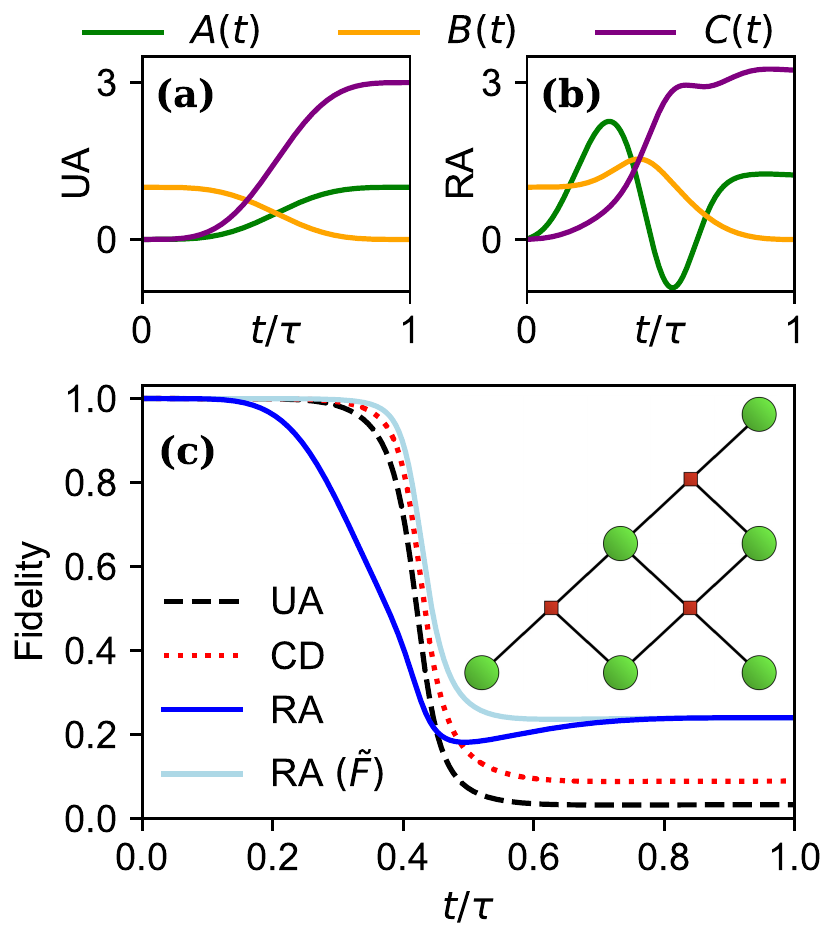}
    \caption{Quantum annealing on the LHZ architecture. (a) and (b) control fields for the UA and RA  protocols. (c) ground state fidelity $F(t)$ for the UA, local CD and RA protocols, and rotated ground state fidelity $\tilde{F}(t)$ for the RA protocol.  The data refers to a system of $N=6$ spins with random couplings $J_j\in [-1,1]$ and protocol's duration $\tau=1$. The inset of (c) shows a cartoon LHZ architecture.}
    \label{fig:lhz_dynamics}
\end{figure}
Quantum annealing (QA)\cite{Kadowaki_Nishimori_PRE1998, Santoro_Car_Sci2002,Hauke_Oliver_RepProgPhys2020} is a strategy to solve classical combinatorial optimization problems on quantum devices. 
Here, we consider the minimization of a generic quadratic unconstrained binary optimization (QUBO) cost function~\cite{Lucas_FrontPhys2014} $f(\boldsymbol{s})=\sum_{\nu<\mu}^n J_{\nu\mu} s_{\nu}s_{\mu}$, which depends on the $n$ binary variables $s_1,\dots, s_n=\pm 1$. 
We adopt the LHZ architecture~\cite{Lechner_Sci2015} to translate the long-range QUBO problem into an experimentally accessible short-ranged spin model, where the physical spins $\PauliSigma_{\nu}^z\PauliSigma_{\mu}^z$ represent the relative configuration of two $s_{\nu}s_{\mu}\to \PauliSigma_{j}^z$. 
The resulting LHZ quantum annealing Hamiltonian reads
\begin{align}
   \Ham_{\mathrm{LHZ}} &= -A(t) \sum_{k=1}^NJ_k\PauliSigma_k^z -B(t)\sum_{k=1}^N\PauliSigma_k^x - C(t)\sum_{l}^{L}\hat{H}_{\squaredots,l} \nonumber
\end{align}
where $N=n(n-1)/2$ is the number of physical qubits and  $L=(n-1)(n-2)/2$ is the number of 4-body constrains $\hat{H}_{\squaredots,l}=\PauliSigma^z_{(l,1)}\PauliSigma^z_{(l,2)}\PauliSigma^z_{(l,3)}\PauliSigma^z_{(l,4)}$ required to implement the parity mapping. The inset of Fig.~\ref{fig:lhz_dynamics} shows the LHZ architecture for $n=4$ (and $N=6$ qubits).
Following Ref.~\cite{Lechner_Sci2015}, we consider the UA schedule $A_0(t)=\lambda$, $B_0(t)=1-\lambda$, $C_0(t)=C_f\lambda$ [see Fig.~\ref{fig:lhz_dynamics}(a)]. Here,  we the final constrains to be $C_f=3$, which ensure that the final ground state encodes the solution of the considered QUBO problem~\cite{Lechner_Sci2015,Lanthaler_NewJPhys2021}. 

We parametrize the auxiliary potentials as $\Qop=-\gamma\sum_{k}J_k\PauliSigma_k^z- \phi \sum_{l}\hat{H}_{\squaredots,l}$ and $\Kop=-\beta\sum_{k}\PauliSigma_k^x$. 
Then, the RA driving functions read $A_{\rmra}=A_0 +\dot{\gamma}$, $C_{\rmra}=C_0+\dot{\phi}$ and $B_{\rmra}=B_0+\beta$. The action for this problem depends on $N$ and the specific disorder implementation. In the \supmat{}we compute the analytical expression of the action and show that its numerical evaluation requires a linear time in the number of qubits $N$. Furthermore, we show in the \supmat{} that similar results hold for conventional quadratic quantum annealing  architectures. 

We benchmark the RA approach on a spin-glass problem with uniform random couplings $J_k\in[-1,1]$. We first analyze a single problem instance of $N=6$ qubits. Figure~\ref{fig:lhz_dynamics}(b) shows the optimal RA control fields obtained by minimizing the action. In Fig.~\ref{fig:lhz_dynamics}(c), we compare the instantaneous ground state fidelity of the UA, the local CD, and the RA protocols on the same problem instance. Despite the disorder, we find a behavior similar to what we observed for the Ising chain. The UA fidelity of $F_{\rmua}(\tau)=4\%$, a local CD fidelity of $F_{\rmlocalcd}(\tau)=8\%$ and $F_{\rmra}(\tau)=18\%$. The RA tracks the rotated ground states with higher fidelity than the UA and local CD protocols. This fact is particularly surprising because the inhomogeneous local CD driving uses $N=6$ additional control fields and does not outperform the RA driving.

\begin{figure}
    \centering
    \includegraphics[width=8.5cm]{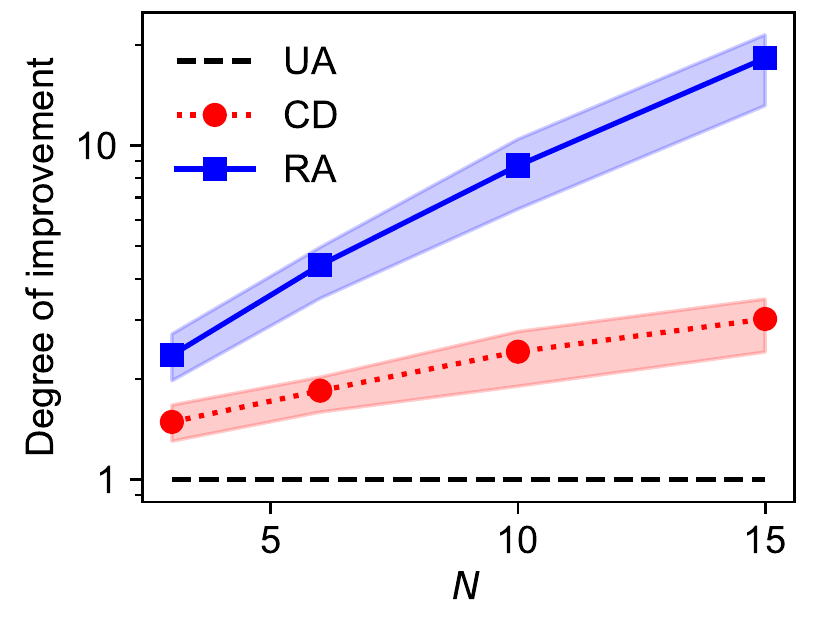}
    \caption{Finite size scaling on the LHZ architecture. relative improvement the local CD $F_{\rmlocalcd}(\tau)/F_{UA}(\tau)$ and for the RA protocol $F_{\rmra}(\tau)/F_{UA}(\tau)$. The points were averaged over $100$ different instances random couplings $J_j\in [-1,1]$. The shaded region represents represents the points between the 25th and 75th percentiles. The protocols duration is $\tau=1$.}
    \label{fig:lhz_scaling}
\end{figure}
Finally, we investigate the finite-size scaling of the protocols' relative performance by considering the relative improvement, which we define as the ratio between the protocol's and the baseline UA final fidelities $\frac{F(\tau)}{F_{\rmqa}(\tau)}$. Figure~\ref{fig:lhz_scaling} shows the relative improvement of the RA and local CD, averaged over an ensemble of $100$ random problems with different qubits numbers $N=4,6,10,15$. Although the limited data points prevent us from identifying scaling forms for the degree of improvement, our numerical results suggest that the features observed for $N=6$ also hold for larger systems. In particular, the relative performance of the RA driving increases with the number of qubits. In the \supmat{}, we examine RA protocols for traditional commercially available quadratic annealing architectures~\cite{Johnson_Rose_Nat2011}, obtaining comparable results.

\mainsection{Conclusion and outlook.}
We have extended the least action formalism of CD drivings to include 
physical rotations of the driving Hamiltonian. The emerging rotated ansatz approach generates alternative variational STAs,  simultaneously optimizing the rotation's generator $\Qop$ and the auxiliary potential $\Kop$. The resulting RA protocols do not require additional control fields and can be readily implemented on analog NISQ devices.

For small instances of spin systems, the RA approach systematically generates experimentally feasible protocols that considerably outperform the unassisted and local CD protocols. Moreover, our scaling analysis suggests that the enhancement persists in larger systems, where RA protocols may aid state preparation and quantum optimization. 

The variational RA approach differs from the optimal control and machine learning methods for schedule optimization considered in Refs.~\cite{Cepaite_Duncan_arXiv2022,Yao_Bukov_PRX2021}, because we can efficiently compute the optimal RA protocol, without having access to a quantum device~\cite{Hartmann_Lechner_NJPhys2019,Susa_Nishimori_PRA2021,Henson_Truscott_PNAS2018}. However, it may be fruitful to investigate the use of optimal control and machine learning techniques to further improve the RA protocol's fidelity. For example, future investigations can use optimal control to select the best initial prior for the variational scheme, as proposed in Ref.~\cite{Cepaite_Duncan_arXiv2022}. 

The extension of the RA approach to include such non-abelian transformations is also an appealing future research direction. Indeed, the unitary transformations that relate fast-forward to counterdiabatic protocols are generally complex and beyond the presented abelian case. We, therefore,  expect that developing variational methods to approximate non-abelian will lead to a more significant enhancement of the protocols' performance.

\mainsection{Acknowledgements}
We thank A. Hartmann, J. Wurtz and C. Dlaska for valuable discussions. This work was supported by the Austrian Science Fund (FWF) through a START grant under Project No. Y1067-N27 and the SFB BeyondC Project No. F7108-N38, and the European Union's Horizon 2020 research and innovation program under grant agreement No. 817482.

\bibliography{biblio}

\end{document}


\title{ Supplementary Material -- Rotated ansatz for approximate counterdiabatic driving}
\author{Glen Bigan Mbeng}
\affiliation{Universit\"at Innsbruck, Technikerstra{\ss}e 21 a, A-6020 Innsbruck, Austria}

\author{Wolfgang Lechner}
\affiliation{Universit\"at Innsbruck, Technikerstra{\ss}e 21 a, A-6020 Innsbruck, Austria}
\affiliation{Parity Quantum Computing GmbH, A-6020 Innsbruck, Austria}

\date{\today}

\maketitle

\section{Limitations of the variational counterdiabatic approach}
This section highlights some limitations of the traditional variational CD approach~\cite{Sels_Polkovnikov_PNAS2017} (without frame rotations). 

We consider the case where the two Hamiltonians $\Ham_a$ and $\Ham_b$ generate the space of
allowed physical operators~\cite{Farhi_Sci2001,Albash_Lidar_RevModPhys2018}. 
Without any further assumption on the form of $\Ham_a$ and $\Ham_b$, we consider the parametric Hamiltonian
\begin{align}\label{eqn:app:H0-2parm_CDlimitations}
    \Ham_0(t) = A_0(\lambda)\Ham_a + B_0(\lambda)\Ham_b\;,
\end{align}
where $A_0(\lambda)$ and $B_0(\lambda)$ are arbitrary real schedule functions, describing the original UA driving. 
Within the space of allowed operators, the most general prametrization for the variational AGP reads
 \begin{equation}\label{eqn:app:Agp-cd-2parm}
    \Agp=\alpha_a  (\lambda) \Ham_a + \alpha_b  (\lambda) \Ham_b\;,
 \end{equation}
where $\alpha_a  (\lambda)$ and $\alpha_b  (\lambda)$ are variational parameters. The CD Hamiltonian is
 \begin{equation}\label{eqn:app:Hcd-2parm}
    \Ham_{\rmcd}(t)=\Ham_0(\lambda) + \dot{\lambda}\Agp=A_{\rmcd}(t)\Ham_a + B_{\rmcd}(t)\Ham_b\;,
 \end{equation}
where $A_{\rmcd}(t) = A_0(t)+\dot{\lambda}\alpha_a(t)$ and $B_{\rmcd}(t) = B_0(t)+\dot{\lambda}\alpha_b(t)$ are the control fields of the CD protocol. Inserting Eq.~\eqref{eqn:app:H0-2parm_CDlimitations} and Eq.~\eqref{eqn:app:Agp-cd-2parm} into  Eq.~\eqref{eqn:CD_action} of the main text, we get the following expression for the action
\begin{align}
\calS (\Agp)
&=\Tr\Big(\Gop_{\lambda}^2\Big) =   
\Tr\left((\partial_\lambda A_0 \Ham_a + \partial_\lambda B_0 \Ham_b - i[A_0\Ham_a + B_0\Ham_b, \alpha_a \Ham_a + \alpha_b\Ham_b])^2\right) \nonumber\\
&= \Tr\Big((\partial_\lambda\Ham_0)^2\Big) + \Tr\Big([\Ham_a,\Ham_b]^2 \Big)(A_0\alpha_b - B_0 \alpha_a)^2\;.\label{eqn:app:action-cd-2parm}
\end{align}
Eq.~\eqref{eqn:app:action-cd-2parm} implies that, the trivial choice $\alpha_a(\lambda)=\alpha_b(\lambda)=0$ is a global minimum of the action. Hence, the  traditional CD variational approach, without frame rotations, returns the trivial result $A_{\rmcd}(t) = A_0(t) $ and $B_{\rmcd}(t) = B_0(t)$, which gives no performance improvement. However, as we showed in the main text, the RA method can overcome this performance limitation.

%

\section{Boundary conditions for the rotated protocols}
 In this section, we show that by choosing a ramp $\lambda(t)$ such that $\dot{\lambda}(0)=\dot{\lambda}(\tau)=0$ we can enforce the time boundary condition, required by RA protocols.

We recall that for the RA driving to result in an alternative STA, we need the time boundary condition, given in Eq.~\eqref{eqn:ra_boundary_conditions} of the main text to hold for the initial and final times $t=0,\tau$. Namely, we must require
\begin{align}\label{eqn:app:ra_boundary_conditions-1}
    |\braket{\epsilon(\lambda)|\nepero^{ -\frac{i}{\hbar} \Qop }|\psi(t)}|&=|\braket{\epsilon(\lambda)|\psi(t)}|
    \qquad \mathrm{for}\; t = 0,\tau 
\end{align}

We now consider the specific case of $\dot{\lambda}\to 0$ for $t=0,\tau$.
We scale the action by the non negative (time-dependent) factor $\dot{\lambda}^2$. The scaled action $\bar{\calS}(\Agp)$ is
\begin{align}
    \bar{\calS}(\Agp) = \dot{\lambda}^2 \calS(\Agp) 
    &= \Big((\dot{\lambda}\partial_\lambda \Hparam - \frac{i}{\hbar} [\Hparam, \dot{\lambda}\Agp])^2\Big) \nonumber\\
    &= \Big((\dot{\lambda}\partial_\lambda \Hparam - \frac{i}{\hbar} [\Hparam, \nepero^{ \frac{i}{\hbar} \Qop }(\Ham_0 + \Kop)\nepero^{-\frac{i}{\hbar} \Qop }])^2\Big)
    \label{eqn:app:ra_boundary_conditions}
\end{align}
For $\dot{\lambda}^2> 0$, the optimal auxiliary potentials also minimize $\bar{\calS}(\Agp)$. Assuming that $\partial_\lambda \Hparam$ is bounded, we get
\begin{align}
     \lim_{\dot{\lambda} \to 0} \bar{\calS}(\Agp) =\frac{1}{\hbar^2}\Big([\Hparam, \nepero^{ \frac{i}{\hbar} \Qop }(\Ham_0 + \Kop)\nepero^{-\frac{i}{\hbar} \Qop }])^2\Big)\;.
    \label{eqn:app:ra_boundary_conditions-biss}
\end{align}
In particular, if $\dot{\lambda}=0$ for $t=0,\tau$,  by taking $\Kop$ and $\Qop$ such that
\begin{align}
    \nepero^{ -\frac{i}{\hbar} \Qop } = \idop, \qquad \mathrm{ and }\; \Kop = \zeroop
\end{align}
we simultaneously minimize $\bar{\calS}(\Agp)$ and satisfy Eq.~\eqref{eqn:app:ra_boundary_conditions-1}.  Hence, we can enforce the time boundary condition by choosing $\lambda(t)$ such that $\dot{\lambda}(0)=\dot{\lambda}(\tau)=0$.

\section{Sequential local optimization of the action}
In the following, we describe the numerical algorithm we use to optimize the RA action and compute the RA driving protocols. 

To simplify the description of the algorithm, we explicitly consider the case where the two Hamiltonians $\Ham_a$ and $\Ham_b$ generate the space of allowed physical operators~\cite{Farhi_Sci2001,Albash_Lidar_RevModPhys2018}. Without any further assumption on the form of $\Ham_a$ and $\Ham_b$, we consider the parametric Hamiltonian
\begin{align}\label{eqn:app:H0-2parm_for_sequential-algorithm}
    \Ham_0(t) = A_0(\lambda)\Ham_a + B_0(\lambda)\Ham_b\;,
\end{align}
where $A_0(\lambda)$ and $B_0(\lambda)$ are arbitrary real schedule functions, describing the original UA driving. Within the space of allowed operators, we parameterize the auxiliary RA potentials as 
\begin{align}
   \Qop&=\gamma\Ham_a\\
   \Kop&=\beta \Ham_b
\end{align}
where $\beta(t)$ and $\gamma(t)$ are variational parameters. 
The corresponding AGP and RA driving Hamiltonian read
\begin{align}
    \dot{\lambda}\Agp
    &=     \nepero^{ \frac{i}{\hbar} \Qop } 
\left(\Ham_0 + \Kop\right)     \nepero^{ -\frac{i}{\hbar} \Qop } 
- \Ham_0 = (B_0 + \beta) \nepero^{ i \gamma \Ham_a } 
\Ham_b    \nepero^{ -i \gamma \Ham_a } 
 - B_0\Ham_b\label{eqn:app:Agp-ra-2parm}\\
    \Hra &= \Hparam+\Kop + \dot{\Qop} = A_{\rmra}(t)\Ham_a + B_{\rmra}(t)\Ham_b \label{eqn:app:Hra-2parm}
\end{align}
where $A_{\rmra}(t) = A_0(t) + \dot{\gamma}(t)$ and $B_{\rmra}(t) = B_0(t)+\beta(t)$ are the control fields of the RA protocol. The expression of the AGP leads to the following action
\begin{align}
\calS (\Agp)
&=\Tr\Big(\Gop_{\lambda}^2\Big) =   
\Tr\left((\partial_\lambda A_0 \Ham_a + \partial_\lambda B_0 \Ham_b - i[A_0\Ham_a + B_0\Ham_b, (B_0 + \beta) \nepero^{ i \gamma \Ham_a } 
\Ham_b    \nepero^{ -i \gamma \Ham_a } 
 - B_0\Ham_b])^2\right) \;. \label{eqn:app:action-ra-2parm}
\end{align}
We observe that the trivial choice $\beta(t)=\gamma(t)=0$ is generally not a global minimum of the action. Hence, to find the functions $\gamma(t)$ and $\beta(t)$, we must explicitly minimize Eq.~\eqref{eqn:app:action-ra-2parm}.

Since $A_{\rmra}(t) = A_0(t) + \dot{\gamma}(t)$, we are specifically looking for differentiable functions and we are not necessarily interested in finding the global optimum of the action for each $t\in[0,\tau]$. We, therefore, use a sequential local optimization algorithm to compute the functions in the interval $t\in[0,\tau]$. We first consider the discrete time problem of finding $\beta^{(m)}=\beta(t^{(m)} )$ and $\gamma^{(m)}=\gamma(t^{(m)} )$, for $t^{(m)} =\frac{m-1}{M}\tau$ and $m=1, \dots M+1$. Here the integer $M$ indicates the number of discrete points considered. Since we are ultimately interested in generating smooth differentiable functions, we can assueme that $\beta^{(m)}$ and  $\gamma^{(m)}$ are close to $\beta^{(m+1)}$ and  $\gamma^{(m+1)}$ respectively. This observation suggests to use sequential local optimizations to solve the discrete problem. Starting from given initial values, eg. $\gamma^{(0)}=\mathbf{0}$, $\beta^{(0)}=\mathbf{0}$, we use the BFGS algorithm~\cite{Nocedal_Wright_book2006} to minimize the action at time $0$, which returns the optimal values $\beta^{(1)}$ and $\gamma^{(1)}$. Then, we proceed iteratively, using $\beta^{(m)}$ and  $\gamma^{(m)}$ as initial guess to find $\beta^{(m+1)}, \gamma^{(m+1)}$. Finally, we interpolate the resulting $M$ points to get the functions $\beta(t)$ and $\gamma(t)$, which we, then, use to obtain the new schedules $A_{\rmra}(t)$ and $B_{\rmra}(t)$. 

The algorithm's generalization to multiple variational parameters case is straightforward. The computation of RA protocols $A_{\rmra}(t)$ and $B_{\rmra}(t)$ does not require spectral information and can run on a classical CPUs. The algorithm's bottleneck is the evaluation of the action, which can still be exponentially hard, as it involves the multiplication of large matrices. In particular, in a system of $N$ spins, the matrix multiplication's complexity scales exponentially with $N$. However,
as reported in the main text, for various physically relevant Hamiltonians,  the complexity of computing RA action is only polynomial in the system size $N$.

\section{Rotated ansatz protocol for a two-level system}\label{app:2level_system}
This section provides additional details on the two-spin system studied in the main text.  Refs.~\cite{Petiziol_PRA2018,Sels_Polkovnikov_PNAS2017} used the two-level nature of the system to compute an analytical expression for the adiabatic gauge potential. Here, we use the same problem to illustrate the RA approach. We use natural units and set $\hbar=1$, we indicate with $\dot{x}(t)$ the time derivative of a given function $x(t)$. 

The parametric Hamiltonian of the two-spin system considered in Eq.~\eqref{eqn:2spin_H0} of the main text is
\begin{equation}\label{eqn:app:H0-2level}
   \Hparam(\lambda) =  h_0(\lambda) \Ham_a + J_0(\lambda)\Ham_b\;,
\end{equation}
where
\begin{align}
   \Ham_a &= -\PauliSigma^z_1 - \PauliSigma^z_2, \qquad \Ham_b = \PauliSigma^x_1\PauliSigma^x_2 +\PauliSigma^z_1\PauliSigma^z_2
\end{align}
generate the space of allowed physical operators. 

We parametrize the auxiliary potentials as 
\begin{align}
   \Qop&=\gamma\Ham_a\\
   \Kop&=\beta \Ham_b
\end{align}
where $\beta(t)$ and $\gamma(t)$ are variational parameters. 
The corresponding RA driving Hamiltonian reads
\begin{align}
    \Hra &= \Hparam+\Kop + \dot{\Qop} = h_{\rmra}(t)\Ham_a + J_{\rmra}(t)\Ham_b\;. \label{eqn:app:Hra-2level}
\end{align}
where $h_{\rmra}(t) = h_0(t) + \dot{\gamma}(t)$ and $J_{\rmra}(t) = J_0(t)+\beta(t)$ are the control fields of the RA protocol. To obtain the optimal values of $\beta$ and $\dot{\gamma}$, we must minimize the RA action.

\subsection{RA adiabatic potential and RA action }

To simplify the computation of the RA Adiabatic gauge potential, we exploit the two-level nature of the system. Indeed, Eq.~\eqref{eqn:app:H0-2level} describes a two-level system where $\Ham(t)$
only couples the states $\ket{\uparrow\uparrow}$ and $\ket{\downarrow\downarrow}$. We, therefore, introduce the operators
\begin{align}
    \hat{s}^z=\frac{1}{4}(\PauliSigma^z_1+\PauliSigma^z_2), \qquad 
    \hat{s}^x=\frac{1}{4}( \PauliSigma^x_1\PauliSigma^x_2 - \PauliSigma^y_1\PauliSigma^y_2), \qquad
    \hat{s}^y =\frac{1}{4}(\PauliSigma^x_1\PauliSigma^y_2 + \PauliSigma^y_1\PauliSigma^x_2), \qquad
     \hat{o} &= \frac{1}{2}(\PauliSigma^x_1\PauliSigma^x_2 +\PauliSigma^y_1\PauliSigma^y_2)+\PauliSigma^z_1\PauliSigma^z_2
    \;
\end{align}
The operators $\hat{s}^x$, $\hat{s}^y$ and $\hat{s}^z$ obey spin algebra, and $\hat{o}$ is such that $\hat{o}\hat{s}^{\nu}=\hat{s}^{\nu} \hat{o}=\hat{s}^{\nu}$ for $\nu=x,y,z$. In particular, in the subspace generated by  $\{\ket{\uparrow\uparrow},\ket{\downarrow\downarrow}\}$, the operators $\hat{s}^x$, $\hat{s}^y$ and $\hat{s}^z$ are representet by spin-$1/2$ matrices, while $\hat{o}$ is the identity matrix. 

Using these operators, the two Hamiltonian terms read $\Ham_a = -4\hat{s}^z$  and $\Ham_b =2\hat{s}^x  + \hat{o}$. Inserting these expressions in the definition of the RA adiabatic gauge potential, we get
\begin{align}
    \dot{\lambda}\Agp
    &=     \nepero^{ \frac{i}{\hbar} \Qop } 
\left(\Ham_0 + \Kop\right)     \nepero^{ -\frac{i}{\hbar} \Qop } 
- \Ham_0 = 2(\beta+J_0 )\nepero^{ -i 4\gamma\hat{s}^z } \hat{s}^x\nepero^{ i 4\gamma\hat{s}^z} +\beta\hat{o} - 2J_0 \hat{s}^x \nonumber\\
     &= \beta\hat{o} + 2(\beta+J_0 )(\cos(4\gamma)\hat{s}^x - \sin(4\gamma)\hat{s}^y) - 2J_0\hat{s}^x \nonumber\\
 &= \beta\hat{o} + 4\alpha_{xx}\hat{s}^x + 4\alpha_{xy}\hat{s}^y \nonumber\\
 &= \beta\hat{o} + \alpha_{xx}\PauliSigma^x_1\PauliSigma^x_2 + \alpha_{yy}\PauliSigma^y_1\PauliSigma^y_2 + \alpha_{xy}(\PauliSigma^x_1\PauliSigma^y_2 + \PauliSigma^y_1\PauliSigma^x_2)\label{eqn:app:Agp-ra-2level}
\end{align}
where we introduced the AGP coefficients 
\begin{align}
    \alpha_{xx} = -\alpha_{yy} = \frac{(\beta+J_0 )\cos(4\gamma) - J_0}{2}\,,\qquad
    \alpha_{xy} = -\frac{(\beta+J_0 )}{2}\sin(4\gamma)\;.
\end{align}
Eq.~\eqref{eqn:app:Agp-ra-2level} shows that the frame rotation introduces in the AGP a new term proportional to $\PauliSigma^x_1\PauliSigma^y_2 + \PauliSigma^y_1\PauliSigma^x_2$. We also compute the operator $\Gop$, obtaining
\begin{align}
\dot{\lambda} \Gop &= \dot{\Ham}_0 +\frac{i}{\hbar}\left[\dot{\lambda}\Agp,\Ham_0\right]\nonumber\\
&=   \dot{J}_0(2\hat{s}^x + \hat{o})- 4\dot{h}_0 \hat{s}^z +i\left[ 4\alpha_{xx}\hat{s}^x + 4\alpha_{xy}\hat{s}^y,2 J_0 \hat{s}^x  - 4h_0\hat{s}^z\right]\nonumber\\
&=  \dot{J}_0\hat{o} + 2(\dot{J}_0+8h_0 \alpha_{xy})\hat{s}^x - 16 h_0\alpha_{xx}\hat{s}^y + 4( 2J_0 \alpha_{xy}  -\dot{h}_0 )\hat{s}^z 
\;.\label{eqn:app:G-ra-2level}
\end{align}
Finally, we insert Eq.\eqref{eqn:app:G-ra-2level} into the definition of the action, leading to 
\begin{align}\label{eqn:2level_action_appendix}
\calS 
&=  \Tr(\Gop^2) = 6\dot{J}^2_0 + 2(\dot{J}_0+8h_0 \alpha_{xy})^2+ 2^7 h^2_0\alpha_{xx}^2 + 2^3( 2 J_0 \alpha_{xy}  -\dot{h}_0 )^2
\end{align}

\subsection{RA action minimization and RA protocols}
\begin{figure}
    \centering
    \includegraphics[width=8cm]{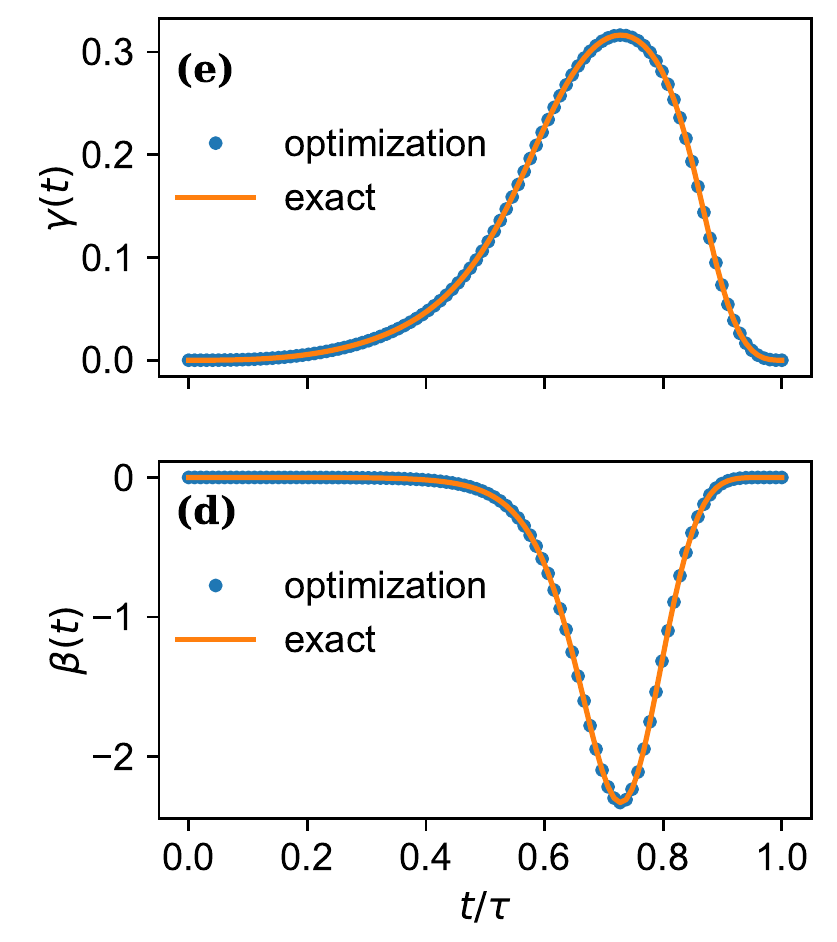}
    \includegraphics[width=8cm]{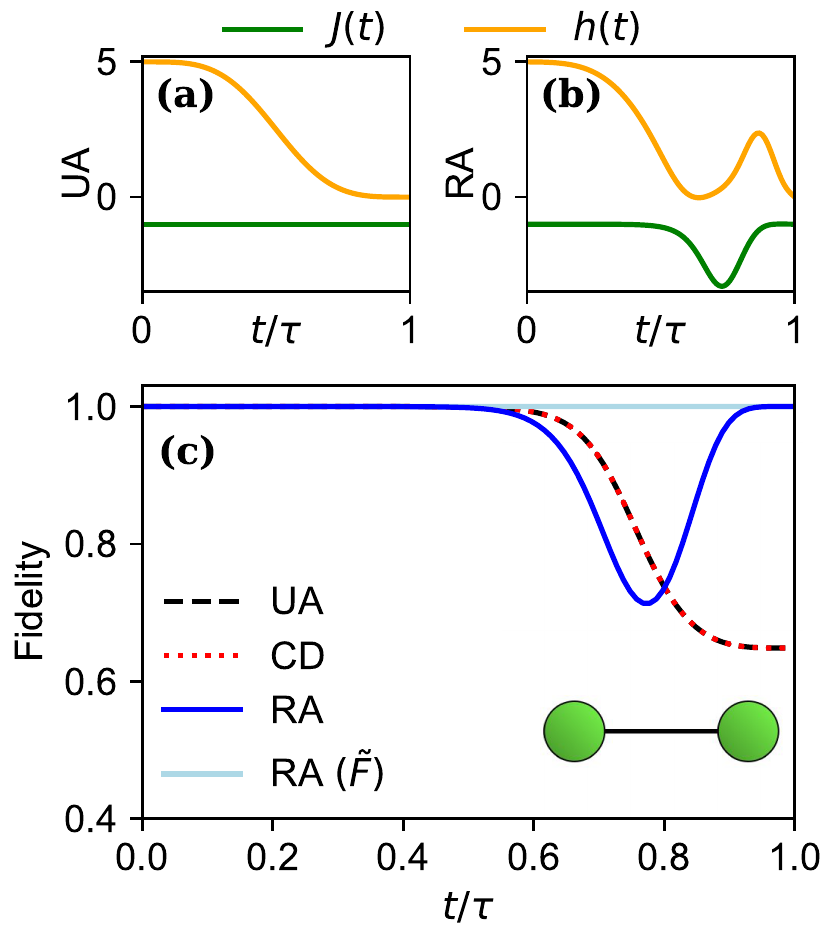}
    \caption{Two-level system. (a) and (b) optimal value of the time-dependent parameter $\gamma(t)$ and $\beta(t)$ respectively. The data was obtained with the sequential local optimization algorithm's results (blue dots) and the exact analytical solutions of Eq.~\eqref{eqn:app:angles-ra-2level} (orange line). (c) and (d) control fields for the UA and RA  protocols. (e) ground state fidelity $F(t)$ for the UA, local CD and RA protocols, and rotated ground state fidelity $\tilde{F}(t)$ for the RA protocol. The inset of (e) shows the two-spin system. The data refers to protocols of total duration $\tau=1$}
    \label{fig:app:2level_system}
\end{figure}
To minimize the action we use the equations $\frac{\partial \calS}{\partial\alpha_{xx}} = \frac{\partial \calS}{\partial\alpha_{xy}}=0$. The quadratic function is minimized at the point
\begin{align}
   \begin{cases}
   \alpha_{xx} &= 0\\
   \alpha_{xy} &= -\frac{1}{2}\varphi_0   
   \end{cases} 
   \implies 
   \begin{cases}
        \cos(4\gamma) &= \frac{J_0}{J_0 + \beta}\\
        \sin(4\gamma) &= \frac{\dot{J}_0 h_0-J_0 \dot{h}_0}{(J_0 + \beta)(4h_0^2 + J_0^2)} =  \frac{\varphi_0}{J_0 + \beta} 
   \end{cases}
\end{align}
where 
\begin{align}
    \varphi_0 = \frac{\dot{J}_0 h_0-J_0 \dot{h}_0}{4h_0^2 + J_0^2}\;. \label{eqn:app:varphi0-ra-2level}
\end{align}
If we assume $J_0+\beta>0$ and $4\gamma \in ( -\frac{\pi}{2}, \frac{\pi}{2})$, we can write 
\begin{align}
    \beta = \sqrt{J_0^2 + \varphi_0^2} - J_0,\qquad
    \gamma  = \frac{1}{4}\arctan \left(\frac{\varphi_0}{J_0}\right)\label{eqn:app:angles-ra-2level}
\end{align}

By substitution Eq.~\eqref{eqn:app:angles-ra-2level}, we get that the optimal RA 
\begin{align}
    \frac{\partial \calS}{\partial\alpha_{xx}} = \frac{\partial \calS}{\partial\alpha_{xy}}=0\implies \Agp =-\frac{1}{2}\varphi_0 (\PauliSigma^x_1\PauliSigma^y_2 + \PauliSigma^y_1\PauliSigma^x_2)+\left(\sqrt{J_0^2 + \varphi_0^2} - J_0\right)\hat{o}\;.
    \label{eqn:app:exact-Agp-2level}
\end{align}
This result is compatible, up to an irrelevant term proportional to $\hat{o}$, with the well-known exact AGP $\Agp^{\mathrm{exact}} = -2\varphi_0 \hat{s}_y$ for two level systems~\cite{Sels_Polkovnikov_PNAS2017}. Hence, the RA Hamiltonian for the two spin problem implements an exact counterdiabatic protocol in the rotated frame, which tracks the instantaneous rotated ground state $\ket{\tilde{\epsilon}_0(\lambda)}$ with perfect unit $\tilde{F}(t)=\big|\braket{\tilde{\epsilon}_0(\lambda)| \psi(t)}\big|^2 = 1$.

The analytical expressions in Eq.~\eqref{eqn:app:angles-ra-2level} are a specific feature of two level systems. We Eq.~\eqref{eqn:app:angles-ra-2level} to benchmark the numerical sequential local optimization algorithm on the two spin problem. As in the main text we start from the UA protocol $J_0(t)=-1$, $h_0(t)=5(1-\lambda)$, of total duration $\tau=1$. In Fig.~\ref{fig:app:2level_system}a and Fig.~\ref{fig:app:2level_system}b we compare the
$M=100$ sequential local optimization data and the analytical expression for $\beta(t)$ and $\gamma(t)$. The results show that the sequential local optimization points agree with the analytical expression. In Fig.~\ref{fig:app:2level_system}c and Fig.~\ref{fig:app:2level_system}d we show the original UA control fields and the resulting RA control fields respectively. In Fig.~\ref{fig:app:2level_system}e we compare the instantaneous ground state's fidelity of an UA and a RA protocol of duration $\tau=1$. The diabatic transitions in the UA dynamics monotonically decreases the fidelity $F_{\rmua}(t)$ (black dashed line) up to the final value $F_{\rmua}(\tau)=0.66$. The RA protocol's fidelity $F_{\rmra}(t)$ (solid blue line), has a non monotonic time dependence. In particular, although for intermediate times $F_{\rmra}(t)$ is lower $F_{\rmua}(t)$, the RA protocol enhances the final fidelity up to $F_{\rmra}(\tau)=1$. Indeed, the implemented RA driving, tracks the instantaneous rotated ground state $\ket{\tilde{\epsilon}_0(\lambda)}$ with perfect fidelity $\tilde{F}_{\rmra}(t)=1$ (light blue dot-dashed line).

\section{Rotated ansatz protocol for the quantum Ising chain \label{app:Ising1D}}
This section provides additional details on the transnational invariant quantum spin chain system studied in the main text (see Eq.~\eqref{eqn:1DIsing_H0}).  Refs.~\cite{Sels_Polkovnikov_PNAS2017} used the least action principle to construct approximate local CD drivings for the spin chain. Here, we use the least action principle to design an approximate RA protocol. We use natural units and set $\hbar=1$, we indicate with $\dot{x}(t)$ the time derivative of a given function $x(t)$. 

The parametric Hamiltonian of the spin chain considered in Eq.~\eqref{eqn:1DIsing_H0} of the main text is
\begin{equation}\label{eqn:app:H0-chain}
     \Ham_{\mathrm{spin-chain}} =   J_0(\lambda)\Ham_a + h_0(\lambda)\Ham_b + b_0(\lambda)\Ham_c \;,
\end{equation}
with 
\begin{equation}
    \Ham_a = -\sum_{j=1}^{N}\PauliSigma_j^z\PauliSigma_{j+1}^z, \qquad 
    \Ham_b = -\sum_{j=1}^{N}\PauliSigma^x_j, \qquad
     \Ham_c = - \sum_{j=1}^{N}\PauliSigma^z_j\;.
\end{equation}

We parametrize the auxiliary potentials as 
\begin{align}
   \Qop&=\gamma\Ham_a + \phi \Ham_c \\
   \Kop&=\beta \Ham_b
\end{align}
where $\beta(t)$, $\gamma(t)$ and $\phi(t)$ are variational parameters. 
The corresponding RA driving Hamiltonian reads
\begin{align}
    \Hra &= \Hparam+\Kop + \dot{\Qop} = J_{\rmra}(t)\Ham_a + h_{\rmra}(t)\Ham_b + b_{\rmra}(t) \Ham_c\label{eqn:app:Hra-chain}
\end{align}
where $J_{\rmra}(t) = J_0(t) + \dot{\gamma}(t)$, $b_{\rmra}(t) = b_0(t) + \dot{\phi}(t)$ and $h_{\rmra}(t) = h_0(t)+\beta(t)$ are the control fields of the RA protocol. To obtain the optimal values of $\beta$, $\dot{\phi}$ and $\dot{\gamma}$, we must minimize the RA action.

\subsection{RA adiabatic potential and RA action }

We consider systems of $N\geq 4$ spins. Using Pauli's matrices algebra to compute the RA adiabatic gauge potential, we get
\begin{align}
    \dot{\lambda}\Agp
    &=     \nepero^{ \frac{i}{\hbar} \Qop } \left(\Ham_0 + \Kop\right) \nepero^{ -\frac{i}{\hbar} \Qop } - \Ham_0
    =- (J_0+\beta)\sum_{j=1}^{N}
    \, \nepero^{ i\gamma(\PauliSigma^z_{j-1}\PauliSigma^z_{j}+\PauliSigma^z_{j}\PauliSigma^z_{j+1}) +i\phi\PauliSigma^z_{j}}\PauliSigma_j^x \nepero^{ -i\gamma(\PauliSigma^z_{j-1}\PauliSigma^z_{j}+\PauliSigma^z_{j}\PauliSigma^z_
    {j+1}) -i\phi\PauliSigma^z_{j}} + h_0\sum_{j=1}^{N} \PauliSigma_j^x\nonumber \\
    &=\alpha_x\sum_{j=1}^{N}\PauliSigma_j^x
    +\alpha_y\sum_{j=1}^{N}\PauliSigma_j^y
    +\alpha_{xz}\sum_{j=1}^{N}(\PauliSigma_{j}^z\PauliSigma_{j+1}^x + \PauliSigma_j^x\PauliSigma_{j+1}^z)
    +\alpha_{yz}\sum_{j=1}^{N}(\PauliSigma_{j}^z\PauliSigma_{j+1}^y + \PauliSigma_j^y\PauliSigma_{j+1}^z)\nonumber\\
    &\hspace{1cm}+\alpha_{zxz}\sum_{j=1}^{N}\PauliSigma_{j-1}^z\PauliSigma_j^x\PauliSigma_{j+1}^z
    +\alpha_{zyz}\sum_{j=1}^{N}\PauliSigma_{j-1}^z\PauliSigma_j^y\PauliSigma_{j+1}^z
    \label{eqn:app:Ara_1DIsing}
\end{align}
where we introduced the coefficients
\begin{align}
    \alpha_x &= h_0-\frac{h_0+\beta}{2}\cos 2\phi\,(\cos 4\gamma+1), \qquad
    \alpha_y =  \frac{h_0+\beta}{2}\sin 2\phi\,(\cos 4\gamma+1), \qquad
    \alpha_{xz} = \frac{h_0+\beta}{2}\sin 2\phi\sin 4 \gamma,\\
     \alpha_{yz} &= \frac{h_0+\beta}{2}\cos 2\phi\sin 4\gamma, \qquad
     \alpha_{zxz} = -\frac{h_0+\beta}{2} \cos 2\phi\, (\cos 4\gamma-1), \qquad
     \alpha_{zyz} = \frac{h_0+\beta}{2} \sin 2\phi\, (\cos 4\gamma-1)    \,.
\end{align}
This RA adiabatic gauge potential explicitly contains terms proportional to $\sum_{j}\PauliSigma_j^y$, $\sum_{j}(\PauliSigma_{j}^z\PauliSigma_{j+1}^y + \PauliSigma_j^y\PauliSigma_{j+1}^z)$ and $\sum_{j}\PauliSigma_{j-1}^z\PauliSigma_j^y\PauliSigma_{j+1}^z$,  which are not available in the original parametric Hamiltonian. 

Finally, for $N\geq 4$, the action reads
\begin{align}
\frac{\calS (\Agp)}{N2^N}
&=\frac{\Tr(\Gop_{\lambda}^2)}{N2^N} =   
\left(4 b^2+8 J^2\right) \alpha _x^2+\left(4 b_0^2+8
   J_0^2\right) \alpha _y^2+\alpha _{\text{xz}}^2 \left(8 b_0^2+8 h_0^2+32 J_0^2\right)+\alpha
   _{\text{yz}}^2 \left(8 b_0^2+32 h_0^2+32 J_0^2\right)\nonumber\\
   &\hspace{1cm}+
   \left(4 b_0^2+8 h_0^2+8 J_0^2\right)\alpha_{\text{zxz}}^2+ \left(4 b_0^2+12
   h_0^2+8 J_0^2\right)\alpha _{\text{zyz}}^2+\alpha _y
   \left(\left(8 b_0^2+16 J_0^2\right) \alpha _x+32 b_0 J_0 \alpha
   _{\text{xz}}+16 J_0^2 \alpha _{\text{zxz}}\right)\nonumber\\
   &\hspace{1cm}+\alpha _x \left(32 b_0 J_0 \alpha _{\text{xz}}+16
   J_0^2 \alpha _{\text{zxz}}\right)+32 b_0 J_0 \alpha _{\text{xz}} \alpha
   _{\text{zxz}}+32 b_0 J_0 \alpha _{\text{yz}} \alpha
   _{\text{zyz}}+\left(8 h_0 \dot{J}_0-8 \dot{h}_0 J_0\right) \alpha
   _{\text{yz}}\nonumber\\
   &\hspace{1cm}+\dot{b}_0^2+\dot{h}_0^2+\dot{J}_0^2\,.
   \label{eqn:app:Sra_ising-chain} 
   \end{align}
Eq.~\eqref{eqn:app:Sra_ising-chain} reveals that, due to the Hamiltonian's translation symmetry and locality, the ratio $\frac{\Srot (\Qop,\Kop)}{N2^N}$ is not $N$ dependent. This implies that the optimal RA coupling functions $J^{\rmra}(t)$, $b^{\rmra}(t)$ and $h^{\rmra}(t)$ are also $N$ independent, and that the RA protocol can be computed for arbitrarily large systems (with $N\to\infty$).

\subsection{RA action minimization and RA protocols}

The numerical minimization of the action Eq.~\eqref{eqn:app:Sra_ising-chain} with respect to $\beta$, $\gamma$ and $\phi$ leads to the results presented in the main text.

\section{Rotated ansatz protocol for quantum annealing}
In this section, we compute the RA action for transverse field quantum annealing~\cite{Hauke_Oliver_RepProgPhys2020} for a system of $N$ $\frac{1}{2}$-spins. 

We can write the parametric Hamiltonian for transverse field quantum annealing as
\begin{align}\label{eqn:TF_QA_H0}
    \Hparam(\lambda) = \Ham_z(\lambda)  -\sum_{j=1}^{N}h_{0,j}(\lambda)\PauliSigma^x_j\,
\end{align}
where $h_{0,j}(\lambda)$ are local control fields of the parametric Hamiltonian and  $\Ham_z(\lambda)$ is a $\lambda$-depend operator, which is diagonal in the computational $z$-basis. Here, and in the following, we say that an operator $\Dop_z$ is diagonal if $[\Dop_z, \PauliSigma^z_j]=0$ for $j=1,2,\dots, N$.

We parameterize the auxiliary potentials as 
\begin{align}
   \Qop &= \Qop_z\\
   \Kop(\beta) &= -\sum_{j=1}^{N}\beta_j\PauliSigma^x_j\;.
\end{align}
where $\beta_j(t)$, for $j=1,2,\dots, N$,  are $N$ real time-dependent variational parameters and $\Qop_z(t)$ is an operator diagonal in the computational $z$-basis. Me make no further assumption on $\Qop_z(t)$. The corresponding RA driving Hamiltonian reads
\begin{align}
    \Hra = \Hparam+\Kop + \dot{\Qop} 
    = \Ham_{\rmra,z}(t) - \sum_{j=1}^{N} h_{\rmra,j}(t)\PauliSigma^x_j\label{eqn:app:Hra-qa}
\end{align}
where $h_{\rmra, j}(t) = h_{0,j}(t) + \beta_{j}(t)$ are local control fields of the RA protocol and $\Ham_{\rmra,z}(t) = \Ham_z(t) + \dot{\Qop}_z(t)$ is the RA time dependent diagonal Hamiltonian. To obtain the optimal values of $\beta_j$ and $\dot{\gamma}$, we must minimize the RA with respect to the optimal values of $\beta^{j}(t)$ and $\Ham_{\rmra,z}(t)$ we must minimize the RA action with respect to the variational parameters.

\subsection{RA adiabatic potential and RA action }

To compute the action, we rely on a convenient decomposition of diagonal operators.
Let $\Dop_z$ be an arbitrary diagonal operator. We define
\begin{align}
   \Dop^{[j]}_z &\equiv \frac{1}{2} \PauliSigma_j^z\Dop_z +i\frac{1}{4}\left(\PauliSigma_j^x\Dop_z\PauliSigma_j^y - \PauliSigma_j^y\Dop_z\PauliSigma_j^x\right)\qquad \mathrm{for}\; j=1, 2\dots, N \\
   \Dop^{[-j]}_z 
   &\equiv \frac{1}{2} \Dop_z +\frac{1}{2}\PauliSigma_j^x\Dop_z\PauliSigma_j^x\qquad\mathrm{for}\; j=1, 2\dots, N\;.
\end{align}
The operators $\Dop^{[j]}_z$ and $\Dop^{[j]}_z$ then satisfy properties
\begin{itemize}
    \item [(i)] $\Dop_z = \Dop_z^{[j]}\PauliSigma_j^z + \Dop_z^{[-j]}$ for  $j=1, 2, \dots, N$ \hspace{1cm}  (desired decomposition)%
    \item [(ii)]  $[ \Dop_{z}^{[\pm j]},\PauliSigma_k^z]=0$ for $k=1, 2, \dots, N$ \hspace{1cm}($\Dop^{[\pm j]}_z$ is are diagonal operators)
    \item [(iii)]  $[ \Dop_{z}^{[\pm j]},\PauliSigma_j^x]=[ \Dop_{z}^{[\pm j]},\PauliSigma_j^y]=[ \Dop_{z}^{[\pm j]},\PauliSigma_j^z]=0$ \hspace{1cm}($\Dop^{[\pm j]}_z$
    do not involve $\PauliSigma_j^z$)
\end{itemize}
The properties (i),(ii) and (iii) additionally imply the following identities
\begin{itemize}
    \item[(iv)]
    $
        [\Dop_z,\PauliSigma_j^x]
        =   i2\Dop_z^{[j]}\PauliSigma_j^y$ and $    [\Dop_z,\PauliSigma_j^y] = -i2\Dop_z^{[j]}\PauliSigma_j^x\;. \label{eqn:[Dz,sigmay]_appendix}   
    $
    %
    %
    \item[(v)] $
        \big(\cos \Dop_z\big)^{[k]} 
        =-\sin\big(\Dop_z^{[k]} \big)\sin\big(\Dop_z^{[-k]}  \big)$ and $
    \big(\sin \Dop_z\big)^{[k]} =\sin\big(\Dop_z^{[k]} \big)\cos\big(\Dop_z^{[-k]}  \big)
    $
    \item[(vi)]
    $\Dop_z^{[j][k]} = \Dop_z^{[k][j]}$ and $\Dop_z^{[j][j]} = \zeroop\;. \label{eqn:D[j][j]_appendix}$
\end{itemize}
The rationale behind the decomposition is to explicitly factor out the $\PauliSigma_j^z$ from the diagonal operator. 
Using (i)-(iv), the RA for the gauge potential (defined in Eq.~\eqref{eqn:ra_AGP_definition} of the main text) reads
\begin{align}
    {\dot{\lambda}}\Agp 
    &=\nepero^{ \frac{i}{\hbar} \Qop } \left(\Hparam+\Kop\right)\nepero^{ -\frac{i}{\hbar} \Qop}
     - \Hparam
    =-\sum_{j=1}^{N}(h_{0,j}+\beta_j)\nepero^{ i\Qop_z }\PauliSigma_j^x\nepero^{ -i \Qop_z}+\sum_{j=1}^{N}h_{0,j}\PauliSigma_j^x\\
    &=-\sum_{j=1}^{N}\tbeta_j \cos \left(2\Qop_z^{[j]}\right)\,\PauliSigma_j^x + \sum_{j=1}^{N}\tbeta_j
    \sin \left(2\Qop_z^{[j]}\right)\,\PauliSigma_j^y +\sum_{j=1}^{N}h_{0,j}\PauliSigma_j^x\label{eqn:APP:Ara_QA}\;,
\end{align}
where we introduced the variables $\tbeta_j(t)=\beta_j(t)+h_{0,j}(t)$.
Similarly, using (iv) and (v), the operator $\Gop$ reads
\begin{align}
\dot{\lambda} \Gop &= \dot{\Ham}_0 +\frac{i}{\hbar}\left[\dot{\lambda}\Agp,\Ham_0\right]\nonumber\\
&=\dot{\Ham}_z - \sum_{j=1}^{N}\dot{h}_{0,j}\PauliSigma^x_j +i\left[\sum_{j=1}^{N}h_{0,j}\PauliSigma_j^x+\sum_{j=1}^{N}\tbeta_j \left(\cos \left(2\Qop_z^{[j]}\right)\PauliSigma_j^x - \sin \left(2\Qop_z^{[j]}\right)\PauliSigma_j^y\right),\Ham_z-\sum_{j=1}^{N}h_{0,j}\PauliSigma^x_j\right]\nonumber\\
&= \dot{\Ham}_z - \sum_{j=1}^{N}\dot{h}_{0,j}\PauliSigma^x_j +2\sum_{j=1}^{N}\left(h_{0,j}\PauliSigma_j^y-\tbeta_j\cos\left(2\Qop_z^{[j]}\right)\PauliSigma_j^y -\tbeta_j \sin \left(2\Qop_z^{[j]}\right)\PauliSigma_j^x\right)\Ham_z^{[j]}\nonumber\\
&\hspace{1cm}+2\sum_{j=1}^{N}\sum_{k=1}^{N}h_{0,j}\tbeta_k \left(\left(\cos \left(2\Qop_z^{[j]}\right)\right)^{[j]}\PauliSigma_j^y\PauliSigma_k^x-\left(\sin \left(2\Qop_z^{[j]}\right)\right)^{[j]}\PauliSigma_j^y\PauliSigma_k^y+\sin \left(2\Qop_z^{[j]}\right)\,\PauliSigma_k^z\right)\nonumber\\
&=\dot{\Ham}_z -2\sum_{j=1}^{N} \tbeta_j h_{0,j} \sin \left(2\Qop_z^{[j]}\right)\nonumber\PauliSigma_j^z
- \sum_{j=1}^{N}\left( 2\tbeta_j \Ham_z^{[j]}\sin \left(2\Qop_z^{[j]}\right)+\dot{h}_{0,j}\right)\PauliSigma^x_j\nonumber\\
&\hspace{1cm}+2\sum_{j=1}^{N}\left(h_{0,j}\Ham_z^{[j]}-\tbeta_j\cos\left(2\Qop_z^{[j]}\right)\Ham_z^{[j]}\right)\PauliSigma_j^y+2\sum_{j=1}^{N}\sum_{k=1}^{N}\tbeta_j h_{0,k}\sin \left(2\Qop_z^{[j][k]}\right)\sin \left(2\Qop_z^{[j][-k]}\right)\PauliSigma_k^y\PauliSigma_j^x\nonumber\\
&\hspace{1cm}+2\sum_{j=1}^{N}\sum_{k=1}^{N}\tbeta_j h_{0,k}\sin \left(2\Qop_z^{[j][k]}\right)\cos \left(2\Qop_z^{[j][-k]}\right)\PauliSigma_k^y\PauliSigma_j^y
\end{align}

Finally, we use properties~(i), (v) and (vi) to compute the RA action
\begin{align}
\calS 
&=  \Tr(\Gop^2) \nonumber\\
&=\Tr\Bigg(\;\left(\dot{\Ham}_z -2\sum_{j=1}^{N} \tbeta_j h_{0,j} \sin \left(2\Qop_z^{[j]}\right)\PauliSigma_j^z\right)^2
+ \sum_{j=1}^{N}\left(\dot{h}_{0,j} +2\tbeta_j \sin \left(2\Qop_z^{[j]}\right)\Ham_z^{{[j]}}\right)^2\nonumber\\
&\hspace{1cm}+4\sum_{j=1}^{N}\left(h_{0,j}\Ham_z^{[j]}-\tbeta_j\cos\left(2\Qop_z^{[j]}\right)\Ham_z^{[j]}\right)^2
+4\sum_{j=1}^{N}\sum_{k=1}^{N}\tbeta_j^2 h_{0,k}^2\sin^2 \left(2\Qop_z^{[j][k]}\right)\sin^2 \left(2\Qop_z^{[j][-k]}\right)\nonumber\\
&\hspace{1cm}+4\sum_{j=1}^{N}\sum_{k=1}^{N}\tbeta_j^2 h_{0,k}^2\sin^2 \left(2\Qop_z^{[j][k]}\right)\cos^2 \left(2\Qop_z^{[j][-k]}\right)\nonumber\\
&\hspace{1cm}+4\sum_{j=1}^{N}\sum_{k=1}^{N}h_{0,j}\tbeta_j h_{0,k}\tbeta_k \sin^2 \left(2\Qop_z^{[j][k]}\right)\cos \left(2\Qop_z^{[j][-k]}\right)\cos \left(2\Qop_z^{[k][-j]}\right)\Bigg)\nonumber\\
&=\Tr\Bigg(\left(\dot{\Ham}_z \right)^2  -4\sum_{j=1}^{N}\tbeta_j h_{0,j} \dot{\Ham}_z^{[j]}\sin \left(2\Qop_z^{[j]}\right)+8\sum_{j=1}^{N}\sum_{k=1}^{j-1} \tbeta_j h_{0,j}\tbeta_k h_{0,k} \sin^2 \left(2\Qop_z^{[j][k]}\right)\cos \left(2\Qop_z^{[j][-k]}\right)\cos \left(2\Qop_z^{[k][-j]}\right)\nonumber\\
&\hspace{1cm}+4\sum_{j=1}^{N} \tbeta_j^2 h_{0,j}^2 \sin^2 \left(2\Qop_z^{[j]}\right) + 4\sum_{j=1}^{N} \dot{h}_{0,j}\tbeta_j \Ham_z^{{[j]}}\sin \left(2\Qop_z^{[j]}\right) + 4\sum_{j=1}^{N} \tbeta^2_j \left(\Ham_z^{{[j]}}\right)^2\sin^2 \left(2\Qop_z^{[j]}\right)\nonumber\\
&\hspace{1cm}+4\sum_{j=1}^{N}\left(\Ham_z^{[j]}\right)^2\bigg(\tbeta_j^2\cos^2\left(2\Qop_z^{[j]}\right)-2h_{0,j}\tbeta_j\cos\left(2\Qop_z^{[j]}\right)+h_{0,j}^2\bigg)+4\sum_{j=1}^{N}\sum_{k=1}^{N}\tbeta_j^2 h_{0,k}^2\sin^2 \left(2\Qop_z^{[j][k]}\right)\nonumber\\
&\hspace{1cm}+4\sum_{j=1}^{N}\sum_{k=1}^{N}h_{0,j}\tbeta_j h_{0,k}\tbeta_k \sin^2 \left(2\Qop_z^{[j][k]}\right)\cos \left(2\Qop_z^{[j][-k]}\right)\cos \left(2\Qop_z^{[k][-j]}\right)\;\Bigg)
\end{align}
After using property (vi) and the identity $1 - \cos X=2\sin^2X$ to rearrange the terms in the sum, the final expression for the action reads 
\begin{align}
\calS 
&= \Tr\Bigg(\dot{\Ham}_z^2 +\sum_{j=1}^{N}\dot{h}_{0,j}^2 +4\sum_{j=1}^{N} \beta_j^2 \left(\Ham_z^{{[j]}}\right)^2+16\sum_{j=1}^{N}h_{0,j}\tbeta_j\left(\Ham_z^{[j]}\right)^2\sin^2\left(\Qop_z^{[j]}\right)+4\sum_{j=1}^{N}h_{0,j}^2 (\beta_j+h_{0,j})^2 \sin^2 \left(2\Qop_z^{[j]}\right)\nonumber\\
&\hspace{1cm}-4\sum_{j=1}^{N}(\beta_j+h_{0,j}) \left(h_{0,j}\dot{\Ham}_z^{[j]} - \dot{h}_{0,j}\Ham_z^{{[j]}}\right) \sin \left(2\Qop_z^{[j]}\right)+4\sum_{j=1}^{N}\sum_{k=1}^{N}(\beta_j+h_{0,j})^2 h_{0,k}^2\sin^2 \left(2\Qop_z^{[j][k]}\right)\nonumber\\
&\hspace{1cm}-\sum_{j=1}^{N}\sum_{k=1}^{N}h_{0,j}h_{0,k}(\beta_j+h_{0,j}) (\beta_k+h_{0,k}) \sin^2 \left(2\Qop_z^{[j][k]}\right)\cos \left(2\Qop_z^{[j][-k]}\right)\cos \left(2\Qop_z^{[k][-j]}\right)\Bigg)
\;.\label{eqn:app:Sra_QA_v1}
\end{align}

\subsection{RA action minimization and RA protocols}

Eq.~\eqref{eqn:app:Sra_QA_v1} is the starting point to compute the RA action in various models. However, a numerical  evaluation of the action involves computing traces of various $2^N\times2^N$ diagonal matrices. To appreciate the complexity of evaluating the action we use the identity $\nepero^{iX}=\cos X + i\sin X$ into Eq.~\eqref{eqn:app:Sra_QA_v1}, which leads to
\begin{align}
\calS 
&=\Re \Tr\Bigg(\dot{\Ham}_z^2 +\sum_{j=1}^{N}\dot{h}_{0,j}^2 +4\sum_{j=1}^{N} \beta_j^2 \left(\Ham_z^{{[j]}}\right)^2+8\sum_{j=1}^{N}h_{0,j}\tbeta_j\left(\Ham_z^{[j]}\right)^2\left(1-\exp\left(i2\Qop_z^{[j]}\right)\right)\nonumber\\
&\hspace{1cm}+2\sum_{j=1}^{N}h_{0,j}^2 (\beta_j+h_{0,j})^2 \left( 1 - \exp \left(i4\Qop_z^{[j]}\right)\right)+i4\sum_{j=1}^{N}(\beta_j+h_{0,j}) \left(h_{0,j}\dot{\Ham}_z^{[j]} - \dot{h}_{0,j}\Ham_z^{{[j]}}\right) \exp \left(i2\Qop_z^{[j]}\right)\nonumber\\
&\hspace{1cm}+2\sum_{j=1}^{N}\sum_{k=1}^{N}(\beta_j+h_{0,j})^2 h_{0,k}^2\left( 1 - \exp \left(i4\Qop_z^{[j][k]}\right)\right)\nonumber\\
&\hspace{1cm}-\sum_{j=1}^{N}\sum_{k=1}^{N}h_{0,j}h_{0,k}(\beta_j+h_{0,j}) (\beta_k+h_{0,k}) \sum_{s,u,v=\pm1}s \exp \left(i2\left((1+s) \Qop_z^{[j][k]}+u \Qop_z^{[j][-k]}+v\Qop_z^{[k][-j]}\right)\right)\Bigg)\;,
\;,\label{eqn:app:Sra_QA_v2}
\end{align}
where $\Re(x)$ indicates the real part of the complex number $x$.

There are at least two relevant cases for which the RA action~\eqref{eqn:app:Sra_QA_v2} can be computed efficiently:
\begin{itemize}
    \item[(a)] quadratic Hamiltonians: $\Ham_z$ and $\Qop_z$ contain only single-spin and two-spin terms.
    \item[(b)] finite range Hamiltonians: both in $\Ham_z$ and in $\Qop_z$ each spin interacts with at most $\ell$ other spins, with $\ell$ being an $N$ independent natural number.
\end{itemize}
The quadratic Hamiltonian realized on DWAVE annealers~\cite{Johnson_Rose_Nat2011} belongs to (a), while the Hamiltonian associated with parity (or LHZ) architectures~\cite{Lechner_Sci2015} belongs to (b). In the following, we will discuss the two cases separately.

\subsubsection{Quadratic architectures}
The relevant Hamiltonian terms for the QUBO annealing schedules are
\begin{equation}
    \Ham_p = -\frac{1}{2}\sum_{j=1}^{N}\sum_{k=1}^{N}J_{jk}\PauliSigma^z_j\PauliSigma_k^z-\sum_{j=1}^{N}J_{j0}\PauliSigma^z_j , \qquad 
    \Ham_x = -\sum_{j=1}^{N}\PauliSigma^x_j,
\end{equation}
where the couplings $J_{jk}$, for $j,k=0,1,\dots,N$ encode the optimization problem. Without loss of generality we assume that $J_{jk}=J_{kj}$ and $J_{jk}=0$.
The parametric QA Hamiltonian for the QUBO problem is
\begin{equation}\label{eqn:app:H0-qubo}
     \Ham_{\mathrm{QUBO}} =   A_0(\lambda)\Ham_p + B_0(\lambda)\Ham_x\;, 
\end{equation}
where $A_0$ and $B_0$ are the QA control fields.
We parametrize the auxiliary fields as $\Qop=\gamma\Ham_p$ and $\Kop=\beta\Ham_x$. Then, the corresponding RA driving Hamiltonian reads
\begin{align}
    \Hra = \Hparam+\Kop + \dot{\Qop} 
    = A_{\rmra}(t)\Ham_p + B_{\rmra}(t)\Ham_x
\end{align}
where $A_{\rmra}(t) = A_0(t) + \dot{\gamma}(t)$ and $B_{\rmra}(t) = B_0(t)+\beta(t)$ are the control fields of the RA protocol.

To compute the RA action, we first substitute $\Qop_z=\gamma(t)\Ham_p $, $ h_{0,j}(t) = B_0(t)$, $\Ham_z(t)=A_0(t)\Ham_p$ and $\beta_j = \beta(t)$ in Eq.~\eqref{eqn:app:Sra_QA_v2}, 
The final expression of the RA action for the QUBO problem reads 
\begin{align}
\calS(\beta,\gamma)
&=2^{N}\Bigg(\frac{1}{2}\dot{A}^2_0\sum_{j=0}^{N}\sum_{k=0}^{N}J_{jk}^2+N\dot{B}^2_0 + 4\beta^2A_0^2\sum_{j=1}^{N}\sum_{k=0}^{N} J_{jk}^2 +8A_0^2B_0(\beta+B_0)\sum_{j=1}^{N}\sum_{k=0}^{N} J_{jk}^2\left(1 - f_j(\gamma)\right)\nonumber\\
&\hspace{1cm} +16A_0^2B_0(\beta+B_0)\sum_{j=1}^{N}\sum_{k=0}^{N}\sum_{l=0}^{k-1}J_{jk}J_{jl}\tan(2\gamma J_{jk})\tan(2\gamma J_{jl})f_j(\gamma)+2 B_0^2(\beta+B_0)^2\sum_{j=1}^{N} \left(1-f_j(2\gamma)\right)\nonumber\\
&\hspace{1cm}-4(\beta+B_0) (B_0\dot{A}_0 -\dot{B}_0A_0)\sum_{j=0}^{N}\sum_{k=0}^{N}J_{jk} \tan(2 J_{jk}\gamma)f_j(\gamma)+4(\beta+B_0)^2B_0^2\sum_{j=1}^{N}\sum_{k=1}^{N}\sin^2 \left(2 J_{jk}\gamma\right)\nonumber\\
&\hspace{1cm}+4(\beta+B_0)^2B_0^2\sum_{j=1}^{N}\sum_{k=1}^{N}\tan^2 \left(2J_{jk}\gamma\right)\left(g_{+,jk}(\gamma)+g_{-,jk}(\gamma)\right)\Bigg)
\;,\label{eqn:app:Sra_qubo-QA}
\end{align}
where we defined the functions $f_{j}(\gamma) = \prod_{m=0}^{N}\cos(2 J_{jm}\gamma)$ and $ g_{\pm,jk}(\gamma) = \prod_{m=0}^{N}\cos(2J_{jm}\gamma\pm2J_{km}\gamma)$, and we used the following identity $\frac{1}{2^N}\Tr\left(\nepero^{ i\sum_{m=1}^{N}\theta_m\PauliSigma_m^z}\right) =  \prod_{m=1}^{N}\cos(\theta_m)$.

Using Eq.~\eqref{eqn:app:Sra_qubo-QA}, the computational complexity of evaluating the RA action is as $\bigO(N^3)$. Thus, computing RA protocol numerically (with a sequential local optimization) will only result in a polynomial overhead in the number of qubits $N$. 

\begin{figure}
    \centering
    \includegraphics[width=8.5cm]{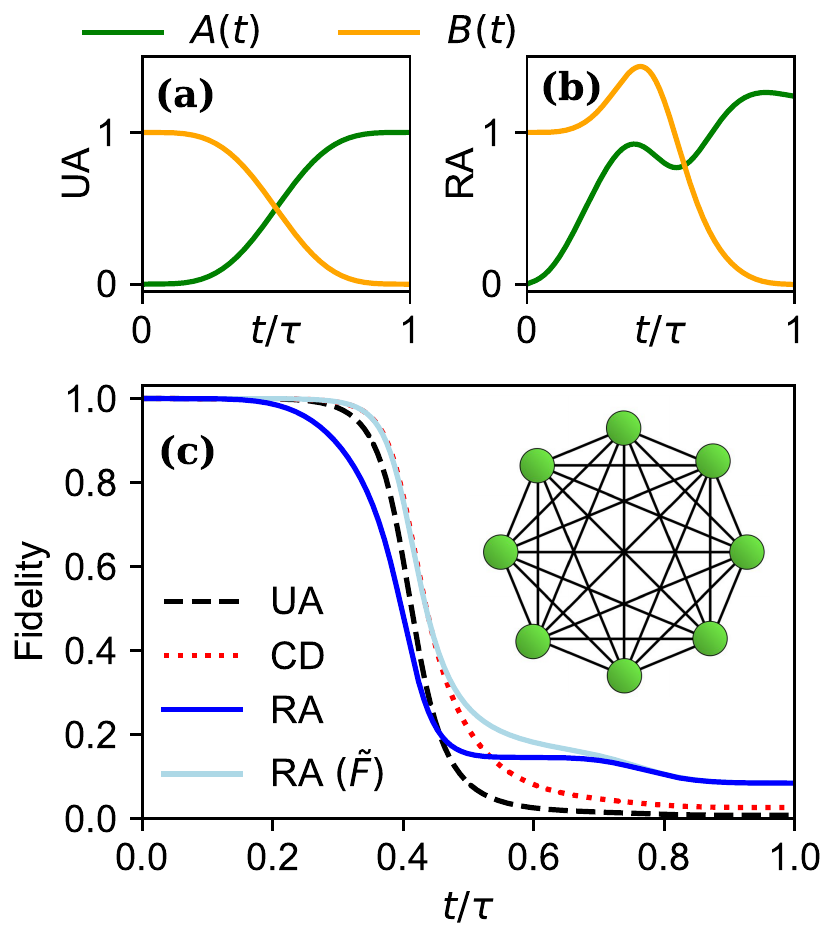}
    \includegraphics[width=8.5cm]{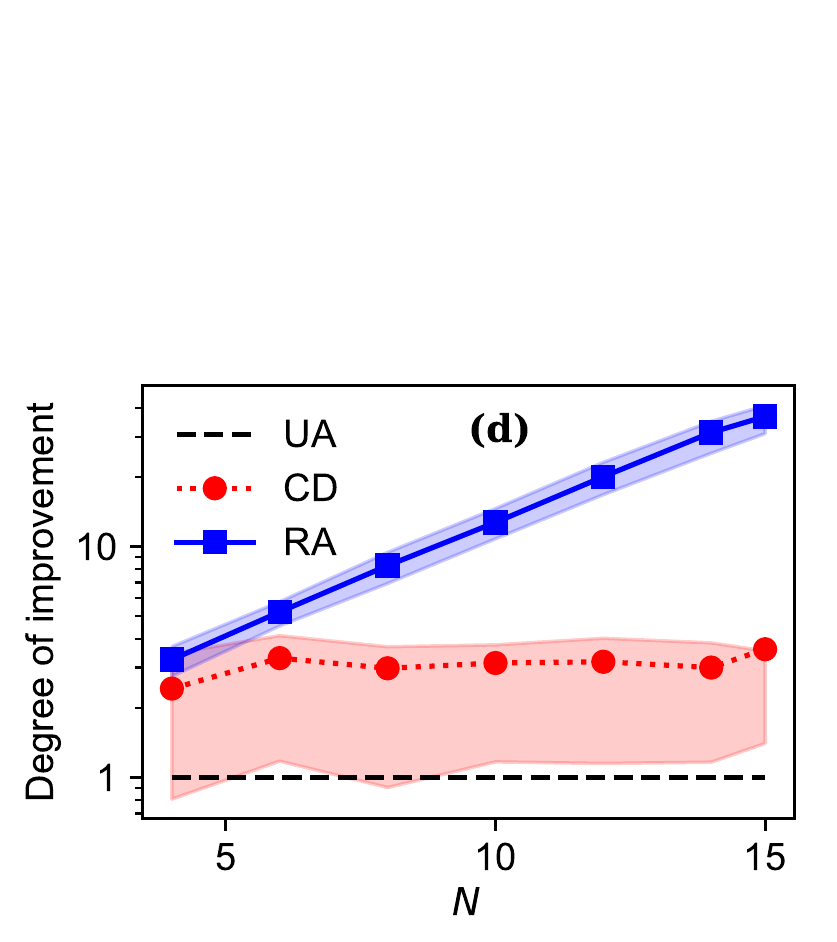}
    \caption{Quantum annealing on the quadratic architecture. (a) and (b) control fields for the UA and RA  protocols. (c) ground state fidelity $F(t)$ for the UA, local CD and RA protocols, and rotated ground state fidelity $\tilde{F}(t)$ for the RA protocol.  The data in (a), (b) and (c) refers to a system of $N=8$ spins with random couplings $J_{jk}\in [-1,1]$, and to protocols of duration $\tau=1$. The inset of (c) shows the fully connected quadratic architecture. (d) shows the the relative improvement the local CD $F_{\rmlocalcd}(\tau)/F_{UA}(\tau)$ and for the RA protocol $F_{\rmra}(\tau)/F_{UA}(\tau)$ as a function of the number of qubits. The points in (d) were averaged over $100$ different instances random couplings $J_{jk}\in [-1,1]$. The shaded region represents the points between the 25th and 75th percentiles. The protocols duration is $\tau=1$.}
    \label{fig:app:qubo_dynamics}
\end{figure}

We benchmark the RA approach on a spin-glass problem with uniform random couplings $J_{jk}\in[-1,1]$ (for $j\neq k$). We start from the UA protocol, shown in Fig.~\ref{fig:app:qubo_dynamics}(a). We first analyze a single problem instance of $N=8$ qubits. Figure~\ref{fig:app:qubo_dynamics}(b) shows the optimal RA control fields obtained by minimizing the action in Eq.~\eqref{eqn:app:Sra_qubo-QA}. In Fig.~\ref{fig:app:qubo_dynamics}(c), we compare the instantaneous ground state's fidelity of the UA, the local CD, and the RA protocols on the same problem instance. We find that the RA tracks the rotated ground states with higher fidelity than the UA and local CD drivings. 
 Fig.~\ref{fig:app:qubo_dynamics}(d) shows the relative improvement of the RA and local CD, averaged over an ensemble of $100$ random problems with different qubits numbers $N=3,4,5,\dots,15$. The data suggest that the features observed for $N=6$ hold for larger systems. As observed in Ref.~\cite{Hartmann_Lechner_PRA2022}, the local CD driving
 shows no scaling advantage over the UA driving. However, the RA protocol provides a scaling advantage over the UA protocol.

\subsubsection{Parity architectures}

The relevant Hamiltonian terms for the LHZ annealing schedules are
\begin{equation}
    \Ham_p = -\sum_{k=1}^NJ_k\PauliSigma_k^z, \qquad \Ham_c=-\sum_{l}^{L}\hat{H}_{\squaredots,l}  \qquad 
    \Ham_x = -\sum_{j=1}^{N}\PauliSigma^x_j,
\end{equation}
where the couplings $J_{j}$, for $j=0,1,\dots,N$ encode the optimization problem. 
The parametric QA Hamiltonian for the LHZ driving is
\begin{equation}\label{eqn:app:H0-lhz}
     \Ham_{\mathrm{LHZ}} =   A_0(\lambda)\Ham_p +  B_0(\lambda)\Ham_x+C_0(t)\Ham_c\;, 
\end{equation}
Where $A_0$,$B_0$ and $C_0$ are the QA control fields.
We parametrize the auxiliary fields as $\Qop=\gamma\Ham_p+\phi\Ham_c$ and $\Kop=\beta\Ham_x$. Then, the corresponding RA driving Hamiltonian reads
\begin{align}
    \Hra = \Hparam+\Kop + \dot{\Qop} 
    = A_{\rmra}(t)\Ham_p + B_{\rmra}(t)\Ham_x+C_{\rmra}(t)\Ham_c
\end{align}
where $A_{\rmra}(t) = A_0(t) + \dot{\gamma}(t)$, $C_{\rmra}(t) = C_0(t) + \dot{\phi}(t)$ and $B_{\rmra}(t) = B_0(t)+\beta(t)$ are the control fields of the RA protocol. 

To compute the RA action, we first substitute $\Qop_z=\gamma(t)\Ham_p +\phi\Ham_c$, $ h_{0,j}(t) = B_0(t)$, $\Ham_z(t)=A_0(t)\Ham_p +C_0(t)\Ham_c$ and $ \beta_j(t) = \beta(t)$ in Eq.~\eqref{eqn:app:Sra_QA_v2}.
The final expression of the RA action for the QUBO problem reads 
\begin{align}
\calS(\beta,\gamma,\phi) 
&=2^N\Bigg(\dot{A}^2_0\sum_{\mu=1}^{N}J_{\mu}^2+N\dot{B}_0^2+L\dot{C}_0^2 +4\left(B_{0}^2+(\beta+B_0)^2\right)\sum_{\mu=1}^{N}  \left(A^2_0J_{\mu}^2+L^{[\mu]} C^2_0\right)\nonumber\\
&\hspace{1cm}-8B_0(\beta+B_0)\sum_{\mu=1}^{N}\cos(2J_\mu \gamma)\left(A_0^2J_\mu^2+C_0^2\left(L^{[\mu]}-L^{[\mu]}(L^{[\mu]}-1)\tan^2(2\phi) \right)\right)\cos^{L^{[\mu]}} (2\phi)\nonumber\\
&\hspace{1cm}+16A_0B_0C_0(\beta+B_0)\sum_{\mu=1}^{N}J_\mu L^{[\mu]}
\sin(2J_\mu \gamma)\tan(2\phi)\cos^{L^{[\mu]}} (2\phi)\nonumber\\
&\hspace{1cm}+2(\beta+B_0)^2 B_{0}^2\sum_{\mu=1}^{N} \left( 1 - \cos(4J_\mu \gamma)\cos^{L^{[\mu]}} (4\phi)\right)\nonumber\\
&\hspace{1cm}-4\left(B_{0}\dot{C}_0 - \dot{B}_{0}C_0\right)(\beta+B_0)\sum_{\mu=1}^{N}  L^{[\mu]}\cos(2J_\mu \gamma)\tan(2\phi)\cos^{L^{[\mu]}} (2\phi)\nonumber\\
&\hspace{1cm}-4\left(B_{0}\dot{A}_0 - \dot{B}_{0}A_0\right)(\beta+B_0)\sum_{\mu=1}^{N}  J_\mu\sin(2J_\mu \gamma)\cos^{L^{[\mu]}} (2\phi) +(\beta+B_0)^2 B_{0}^2\sum_{\langle\mu,\nu\rangle}\left( 1 - \cos^{L^{[\mu][\nu]}} (4\phi)\right)\nonumber\\
&\hspace{1cm}+2B_{0}^2(\beta+B_0)^2\sum_{\langle\mu,\nu\rangle} \cos(2J_\mu\gamma+2J_\nu \gamma)  \left(1-\cos^{L^{[\mu][\nu]}} (4\phi)\right)( \cos 2\phi)^{L^{[\mu][-\nu]}+L^{[\nu][-\mu]}}\Bigg)\;,
\label{eqn:app:Sra_lhz-QA}
\end{align}
Here $L$ is the total number of constrain terms, $L^{[\mu]}$ is the number of constrain terms involving site $\mu$, $L^{[\mu][\nu]}$ is the number of constrain terms involving both sites $\mu$ and $\nu$, and $L^{[\mu][-\nu]}$ is the number of constrain terms involving site $\mu$ but not site $\nu$. Consequently, natural numbers $L$, $L^{[\mu]}$ and $L^{[\mu][\nu]}$ depend only on the architecture, not on the problem instance.  The notation $\sum_{\langle\mu,\nu\rangle}$ indicates a sum over all pairs of interacting spins in the LHZ architecture. To obtain Eq.~\eqref{eqn:app:Sra_lhz-QA}, we used the identity 
\begin{equation}
    \frac{1}{2^N}\Tr\left( \nepero^{i\sum_{l=1}^L\theta_l\hat{H}_{\squaredots,l}^{[\mu]} }\right) =  \prod_{l \in \mathcal{C}_\mu}{\cos \theta_l}\;,
\end{equation}
where $\mathcal{C}_\mu$ is the set of constrains involving the $\PauliSigma^z_\mu$.

Using Eq.~\eqref{eqn:app:Sra_lhz-QA}, the computational complexity of computing the RA action for the LHZ architecture is $\bigO(N)$. Thus, computing RA protocol numerically (with a sequential local optimization) will only result in a polynomial overhead in the number of qubits $N$. 

The numerical minimization of the action given in Eq.~\eqref{eqn:app:Sra_lhz-QA}, for a spin-glass problem leads to the results presented in the main text.

\bibliography{biblio}